\documentclass[useAMS,usenatbib]{mn2e}
\def\aap{AA}
\def\prl{Phys. Rev. Lett.}

\def\apjl{ApJL}
\def\na{NewA}

\def\mnras{MNRAS}
\def\apj{ApJ}
\def\apjs{ApJS}

\def\prd{PhysRevD}
\def\jcap{JCAP}
\def\physrep{Phys. Rept.}
\def\araa{Annual Review of Astronomy and Astrophysics}

\def\dgemail{gilman@astro.utoronto.ca}

\usepackage[titletoc]{appendix}
\usepackage{graphicx}
\usepackage{float}
\usepackage{amssymb}
\usepackage{amsfonts}
\usepackage{amsmath} 
\usepackage{color}
\usepackage{wrapfig}
\usepackage{hyperref}
\usepackage{algorithm}
\usepackage{algpseudocode}
\usepackage{physics}

\usepackage{natbib}

\def\msun{{M_{\odot}}}

\def\dlos{{\delta_{\rm{los}}}}

\title[probing SIDM with strong lensing]{Strong lensing signatures of self-interacting dark matter in low-mass halos} 
\author[Gilman et al.]{\parbox{\textwidth}{
		Daniel Gilman$^{1,2}$\thanks{\dgemail},
		Jo Bovy$^{1}$, 
		Tommaso Treu$^{2}$,
		Anna Nierenberg$^{3}$,
		Simon Birrer$^{4,5}$,
		Andrew Benson$^{6}$,
		Omid Sameie$^{7}$
	} \\
	\\
	\\
	\parbox{\textwidth}{
		$^{1}$Department of Astronomy and Astrophysics, University of Toronto, 50 St. George Street, Toronto, ON, M5S 3H4, Canada\\
		$^{2}$Department of Physics and Astronomy, University of California,
		Los Angeles, CA 90095, USA\\
		$^{3}$Department of Physics, University of California Merced, 5200 North Lake Rd. Merced, CA 95343\\
		$^{4}$Department of Physics, Stanford University, 382 Via Pueblo Mall, Stanford, CA 94305, USA\\
		$^{5}$Kavli Institute for Particle Astrophysics $\&$ Cosmology, P. O. Box 2450, Stanford University, Stanford, CA 94305, USA\\
		$^{6}$Carnegie Observatories, 813 Santa Barbara Street, Pasadena, CA 91101, USA\\
		$^{7}$Department of Astronomy, The University of Texas at Austin, 2515 Speedway, Stop C1400, Austin, TX 78712 USA\\
	}
}
\begin{document}
	
	\voffset-.6in
	
	\date{Accepted . Received }
	
	\pagerange{\pageref{firstpage}--\pageref{lastpage}} 
	
	\maketitle	
	\label{firstpage}
	\begin{abstract}
		Core formation and runaway core collapse in models with self-interacting dark matter (SIDM) significantly alter the central density profiles of collapsed halos. Using a forward modeling inference framework with simulated datasets, we demonstrate that flux ratios in quadruple image strong gravitational lenses can detect the unique structural properties of SIDM halos, and statistically constrain the amplitude and velocity dependence of the interaction cross section in halos with masses between $10^6 - 10^{10} M_{\odot}$. Measurements on these scales probe self-interactions at velocities below $30 \ \rm{km} \ \rm{s^{-1}}$, a relatively unexplored regime of parameter space, complimenting constraints at higher velocities from galaxies and clusters. We cast constraints on the amplitude and velocity dependence of the interaction cross section in terms of $\sigma_{20}$, the cross section amplitude at $20 \ \rm{km} \ \rm{s^{-1}}$. With 50 lenses, a sample size available in the near future, and flux ratios measured from spatially compact mid-IR emission around the background quasar, we forecast $\sigma_{20} < 11-23 \ \rm{cm^2} \rm{g^{-1}}$ at $95 \%$ CI, depending on the amplitude of the subhalo mass function, and assuming cold dark matter (CDM). Alternatively, if $\sigma_{20} = 19.2 \ \rm{cm^2}\rm{g^{-1}}$ we can rule out CDM with a likelihood ratio of 20:1, assuming an amplitude of the subhalo mass function that results from doubly-efficient tidal disruption in the Milky Way relative to massive elliptical galaxies. These results demonstrate that strong lensing of compact, unresolved sources can constrain SIDM structure on sub-galactic scales across cosmological distances, and the evolution of SIDM density profiles over several Gyr of cosmic time. 
	\end{abstract}
	
	\section{Introduction}
	Self-interacting dark matter (SIDM) refers to a class of theories in which dark matter particles can exchange momentum and energy through an interaction weak enough to leave the large scale structure of the Universe unchanged relative to cold dark matter (CDM), but strong enough to alter the internal structure of collapsed halos. Inside halos, the high density of dark matter particles facilitates self-interactions that efficiently conduct heat from the halo's outskirts into the center, supporting the formation of a constant density core \citep{SpergelSteinhardt2000,Burkert2000,Rocha++13,Zavala++13,Nishikawa++20,Nadler++20,Sameie++21}. Over time, the heat reservoir supplied by the outer halo diminishes and the halo experiences `gravothermal catastrophe', or runaway collapse onto itself \citep{LyndenBellandWood68,Balberg++02}. Core formation and collapse result in a rich diversity of structure formation outcomes across many decades in halo mass \citep{Dooley++16,Robles++17,Sameie++18,TulinYu18,Kahlhoefer++19,Zavala++19,Nishikawa++20,Turner++20,Yang++21}. In turn, structure formation measurements over a broad range of halo mass scales constrain the interaction cross section \citep{Kaplinghat++16}.
	
	The form of the interaction cross section can take on a variety of functional forms depending on the the type of interaction and the masses of the interacting species \citep[e.g.][]{Tulin++13b,Colquhoun++20}. To use a concrete example, a weak, long range interaction scales as $v^{-4}$ at high speeds, strongly suppressing the efficiency of self-interactions at relative velocities of $\sim 1000 \ \rm{km}\ \rm{s^{-1}}$ inside galaxy clusters and thus evading stringent constraints on the amplitude of the cross section on these scales \citep{Sand++08,Peter++13,Newman++15,Harvey++15,Kim++17,Banerjee++20,Sagunski++21}. On the other hand, particles inside dwarf galaxies move at typical speeds of $v \sim 50 \ \rm{km} \ \rm{s^{-1}}$, and the larger cross section on these velocity scales drives the formation of central density cores that may eventually collapse. The alterations to the internal structure of dark matter halos with SIDM potentially alleviates tension \citep{Boylan-Kolchin++11,BullockBK17} between the predictions of CDM on small scales and the inferred properties of dwarf galaxies \citep{Valli++18,Kaplinghat++19,Kahlhoefer++19}. 
	
	The possibility that SIDM could resolve the small-scale challenges to the CDM paradigm motivates analyses of SIDM on sub-galactic structure. A popular strategy for testing the viability of SIDM models involves comparing the density profiles of dwarf galaxies inferred from their stellar dynamics with simulations of structure formation \citep{Rocha++13,Zavala++13,Elbert++15,Ren++19,Correa21}. While this approach can in principle differentiate between SIDM and CDM, its limitations stem from reliance on baryonic matter as a tracer for unobservable dark matter. First, the low surface brightness of the smallest dwarf galaxies precludes inferences of their density profiles through stellar dynamics, effectively imposing a minimum halo mass, and hence a minimum velocity scale, where stellar dynamics can constrain the SIDM cross section. Second, baryonic processes couple to the density profile of the dark matter by changing the gravitational potential of a halo \citep[e.g.][]{PontzenGovernato12,Kaplinghat++14,Fry++15,Creasey++17,Sameie++18,Fitts++18,Read++18,Despali++19,Kaplinghat++20}. Thus, the luminous matter required to infer the underlying dark matter density profile can itself alter the dark matter density profile, a complication that obscures the impact of dark matter physics on the observable features of dwarf galaxies. 
	
	Over the past two decades, strong gravitational lensing by galaxies emerged as a powerful tool to constrain dark matter physics on sub-galactic scales, below $10^{10} \msun$ \citep{D+K02,Veg++14,Nierenberg++14,Hezaveh++16,Nierenberg++17,Birrer++17,Hsueh++20,Gilman++20a,Gilman++20b}. Strong lensing circumvents the use of luminous matter as a tracer for dark matter density profiles, and provides perhaps the only means of probing structure in completely dark halos devoid of stars across cosmological distance. While strong lensing has previously constrained halo density profiles and self-interacting dark matter on cluster scales \citep{Meneghetti++01,Sand++08,Newman++15,Andrade++20,Vega-Ferrero++21,Yang++21}, no framework exists to put galaxy-scale strong lensing observables in touch with the predictions of SIDM. 
	
	Recently, \citet{Minor++20} reported that the dark subhalo detected in the lensed arc of the strong lens system SDSSJ0946+1006 \citep{Vegetti++10} has a central density at least an order of magnitude greater than predicted by CDM. \citet{Minor++20} speculated that a core collapsed subhalo could possess a central density high enough to explain this observation. A confirmed detection of a single core collapsed halo would imply the existence of an entire population of similar objects. As strong gravitational lensing directly constrains halo concentrations \citep{Gilman++20b,Minor++20b} and hence the internal structure of halos, the unique structural features predicted in SIDM models may produce a statistically detectable signal in the flux ratios of quadruply imaged quasars (quads). Flux ratios from quads probe halo mass scales down to $10^6 \msun$, depending on the size of the lensed background source, corresponding to constraints on the cross section on velocity scales below $30 \  \rm{km} \ \rm{s^{-1}}$. Quads can therefore probe the cross section at lower velocities than galaxy clusters or Local Group dwarf galaxies through a population-level analysis of dark matter substructure over several decades in halo mass. 
	
	In this work, we use the inference framework developed and tested by \citet{Gilman++18,Gilman++19} to build physical intuition for how flux ratios can constrain models of SIDM, and to forecast constraints from gravitational lensing on self-interacting dark matter. We construct a physical model linking the self-interaction cross section to structure formation in halos with masses between $10^6- 10^{10} \msun$, accounting for both core formation and core collapse, and use the model together with the inference pipeline to simultaneously constrain the velocity dependence and amplitude of the cross section. We assume a functional form for the interaction cross section with a velocity dependence that falls as $v^{4}$ at high velocities in order to remain consistent with constraints on the amplitude at cluster scales, while permitting scattering in low-mass halos efficient enough to change their density profiles.  
	
	This paper is organized as follows: Section \ref{sec:structuremodels} describes the analytic model for the cross section we use to predict halo density profiles, and how we compute cored and core collapsed density profiles for a given cross section. Section \ref{sec:lensingbysidm} discusses the impact of SIDM density profiles on lensing observables, including their deflection angles and magnification cross section. In Section \ref{sec:setupandpriors}, we detail the setup and assumptions built into simulations that we use to forecast constraints on the SIDM cross section using a sample of strong lenses. We present the results of these simulations in Section \ref{sec:results}, and give concluding remarks in Section \ref{sec:conclusions}. 
	
	We assume values for cosmological parameters for flat $\Lambda$CDM measured by WMAP9 \citep{WMAP9cosmo}. We perform lensing computations using the open source gravitational lensing software {\tt{lenstronomy}}\footnote{https://github.com/sibirrer/lenstronomy} \citep{BirrerAmara18,Birrer++21}, and generate subhalo and line of sight halo populations for lensing with the open source software {\tt{pyHalo}}\footnote{https://github.com/dangilman/pyHalo}. We use the halo mass definition of $M_{200}$ with respect to the critical density of the Universe at the halo redshift. 
	
	\section{The density profiles of self-interacting dark matter halos}
	\label{sec:structuremodels}
	
	In this section we describe how we compute the halo density profiles with a velocity-dependent interaction cross section. We begin in Section \ref{ssec:crosssectionmodel} by describing the form of the self-interaction cross section we use in our simulations. Sections \ref{ssec:coredhalos} and \ref{ssec:corecollapse} describe how we implement cored and core collapsed halo density profiles, respectively, as a function of the parameters describing the interaction cross section. 
	
	\begin{figure}
		\includegraphics[clip,trim=0.2cm 0cm 0.5cm
		0.5cm,width=.48\textwidth,keepaspectratio]{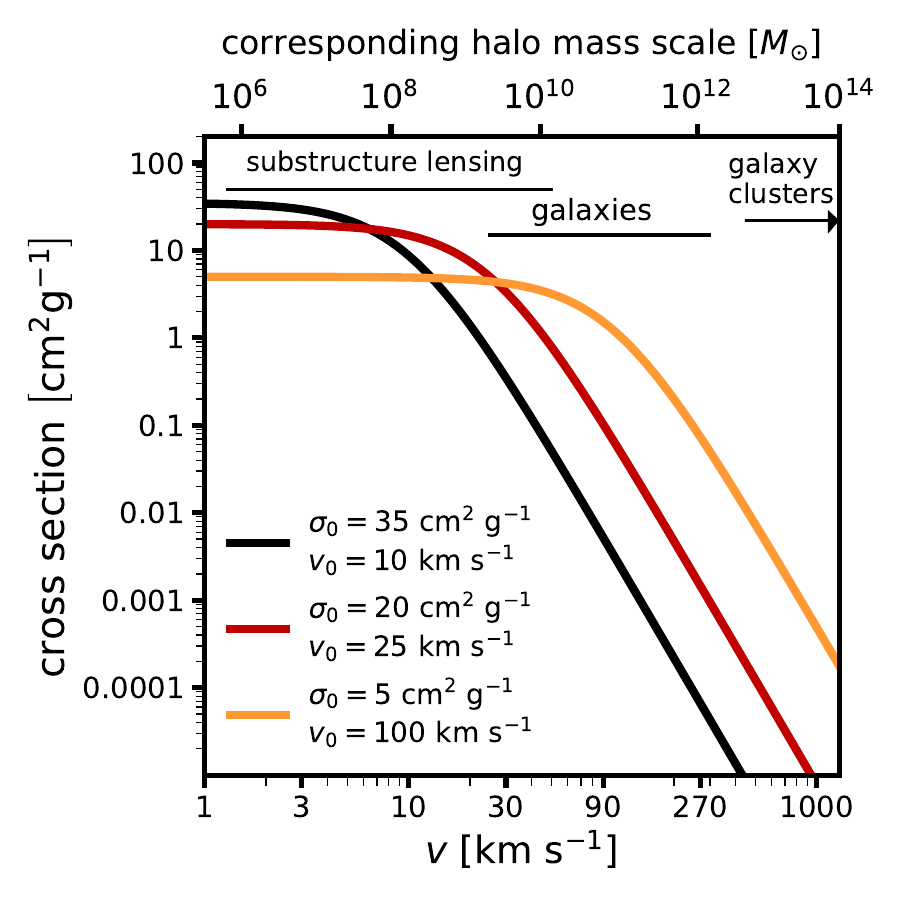}
		\caption{\label{fig:cross_section} The interaction cross section of Equation \ref{eqn:crossec} for three different parameter combinations as a function of relative velocity (lower x-axis), and a corresponding mass scale defined as the halo mass with a central velocity dispersion equal to $v$ inside the scale radius (upper x-axis). The black horizontal lines indicate the scales where substructure lensing (this work), galaxies, and galaxy clusters, can constrain the cross section. The form of the interaction cross section allows for extremely efficient scattering at low velocities while evading constraints on the cross section at higher velocity scales from galaxies and galaxy clusters.}
	\end{figure}
	
	\begin{figure}
		\includegraphics[clip,trim=0cm 0cm 0cm
		0cm,width=.48\textwidth,keepaspectratio]{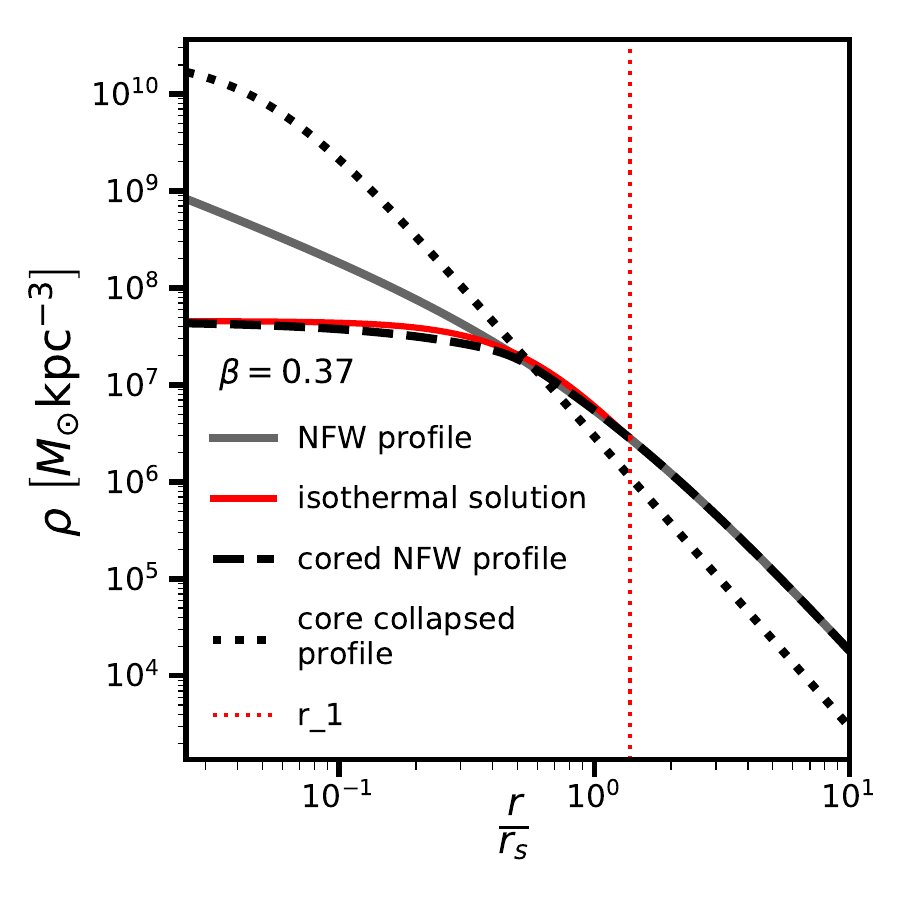}
		\caption{\label{fig:densityprof} An illustration of the possible halo density profiles in CDM and SIDM. The gray curve shows the NFW profile of a $5 \times 10^7 \msun$ halo. The red curve shows the exact solution to Equation \ref{eqn:diffeq} for a cored SIDM halo (red) with cross section parameterized by Equation \ref{eqn:crossec} with $\sigma_0 = 30 \ \rm{cm^2}\rm{g^{-1}}$ and $v_0 = 30 \ \rm{km} \ \rm{s^{-1}}$, resulting in a core radius $\beta \equiv \frac{r_c}{r_s} = 0.37$. We indicate the radius $r_1$ as a vertical red dotted line. The black dashed curve shows our analytic approximation to the exact cored solution, which we parameterize as a cored, truncated NFW profile (Equation \ref{eqn:denistyprof}). Finally, the dashed black curve shows the core collapsed density profile the CDM halo would obtain if it undergoes core collapse using the model described in Section \ref{ssec:corecollapse}.}
	\end{figure}
	
	\subsection{The interaction cross section}
	\label{ssec:crosssectionmodel}
	The differential scattering cross section $\frac{d \sigma}{d \Omega}$ can be computed for any interaction potential by solving the Schr{\"o}dinger equation and using partial wave analysis \citep{Tulin++13b,Colquhoun++20}. In practice, analyses often invoke a proxy for $\frac{d \sigma}{d \Omega}$, labeled $\sigma$. Common choices for $\sigma$ include the momentum transfer cross section, or alternatively, the viscosity cross section, quantities that differ through the integration over the angular dependence of $\frac{d \sigma}{d \Omega}$. We assume a fairly generic form for $\sigma$ parameterized as 
	\begin{equation}
	\label{eqn:crossec}
	\sigma \left(\sigma_0, v_0, v\right) = \sigma_0 \left(1 + \frac{v^2}{v_0^2}\right)^{-2},
	\end{equation}
	which could be interpreted as either the momentum or viscosity cross section. In Equation \ref{eqn:crossec}, $v$ is the relative velocity between particles, and $v_0$ and $\sigma_0$ determine the shape and amplitude of the cross section as a function of relative velocity, respectively. This functional form has similar properties to the model considered by \citet{Ibe++10} and \citet{Nadler++20}, and corresponds to a weak, long range interaction with a velocity dependence that approaches the classical result $\sigma\left(v\right) \propto v^{-4}$ at high $v$. The velocity dependence allows the model to evade stringent constraints on the cross section from galaxy clusters at $v \sim 1000 \rm{km} \ \rm{s^{-1}}$ \citep{Peter++13,Newman++15,Robertson++19,Harvey++15,Kim++17,Andrade++20,Banerjee++20,Sagunski++21}, while allowing for efficient scattering $\mathcal{O}\left(10\right) \ \rm{cm^2} \ \rm{g^{-1}}$ at the scale of dwarf galaxies. Throughout the rest of the text, for brevity we will drop the $\sigma_0$ and $v_0$ dependence in the cross section and simply write $\sigma \left(v\right)$.
	
	Using the terminology of \citet{Kaplinghat++16}, dark matter halos with different masses act as particle colliders with different beam energies, probing the cross section at different velocities. We illustrate this concept in Figure \ref{fig:cross_section}. The colored curves show the cross section in Equation \ref{eqn:crossec} for three different combinations of $\sigma_0$ and $v_0$ as a function of the relative velocity (lower x-axis). The upper x-axis shows a corresponding halo mass scale, which we define as the NFW halo mass with a central velocity dispersion equal to $v$. As Figure \ref{fig:cross_section} illustrates, an interaction cross section that satisfies constraints $\sigma < 10^{-3} \rm{cm^2} \ \rm{g^{-1}}$ at cluster scales could easily reach $10 \ \rm{cm^2} \rm{g^{-1}}$ on the mass scales probed by substructure lensing. A cross section with this amplitude should produce distinct structural features in the halos that perturb strongly lensed images. Specifically, self-interactions should produce either constant density cores, or steep central cusps in core collapsed halos. 
	
	\subsection{Cored SIDM halos}
	\label{ssec:coredhalos}
	\begin{figure}
		\includegraphics[clip,trim=0cm 0cm 0cm
		0cm,width=.48\textwidth,keepaspectratio]{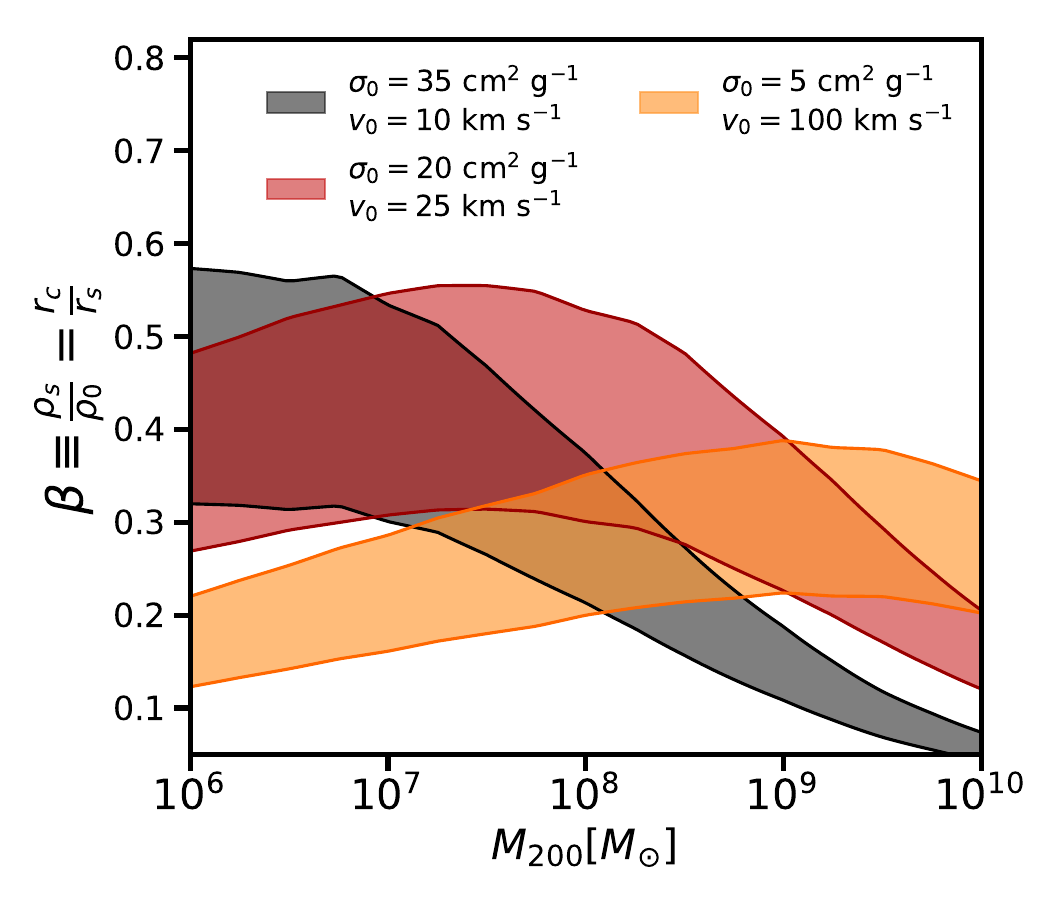}
		\caption{\label{fig:coresizes} The core size in units of the scale radius as a function of halo mass. The three colors show the core size predicted by the model for core formation described in Section \ref{ssec:coredhalos} for the same cross sections shown in Figure \ref{fig:cross_section}. The width of each band comes from the scatter in the mass-concentration relation. The velocity dependence of the cross section $v_0$ alters the slope of the $\beta - M_{200}$ relation, while the normalization $\sigma_0$ rescales the core size at all halo masses.}
	\end{figure}
	
	\begin{figure}
		\includegraphics[clip,trim=0cm 0cm 0cm
		0cm,width=.48\textwidth,keepaspectratio]{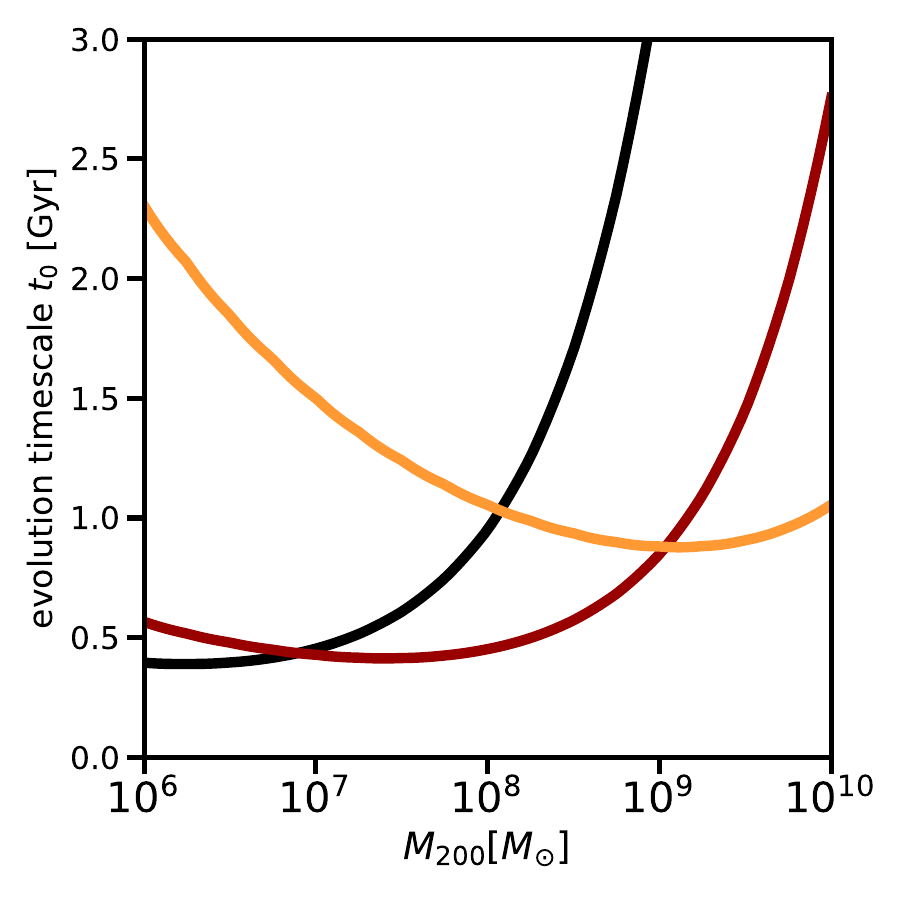}
		\caption{\label{fig:tchalomass} The structural evolution timescale $t_0$ for SIDM halos as a function of halo mass computed with Equation \ref{eqn:collapsetimescale}. The three curves show $t_0$ for the same cross section models shown in Figures \ref{fig:cross_section} and \ref{fig:coresizes}. The velocity dependence of the cross section determines the halo mass where $t_0$ reaches a minimum, and therefore at what halo mass the probability of core collapse reaches a maximum.}
	\end{figure}	
	\begin{figure}
		\includegraphics[clip,trim=0cm 0cm 0cm
		0cm,width=.48\textwidth,keepaspectratio]{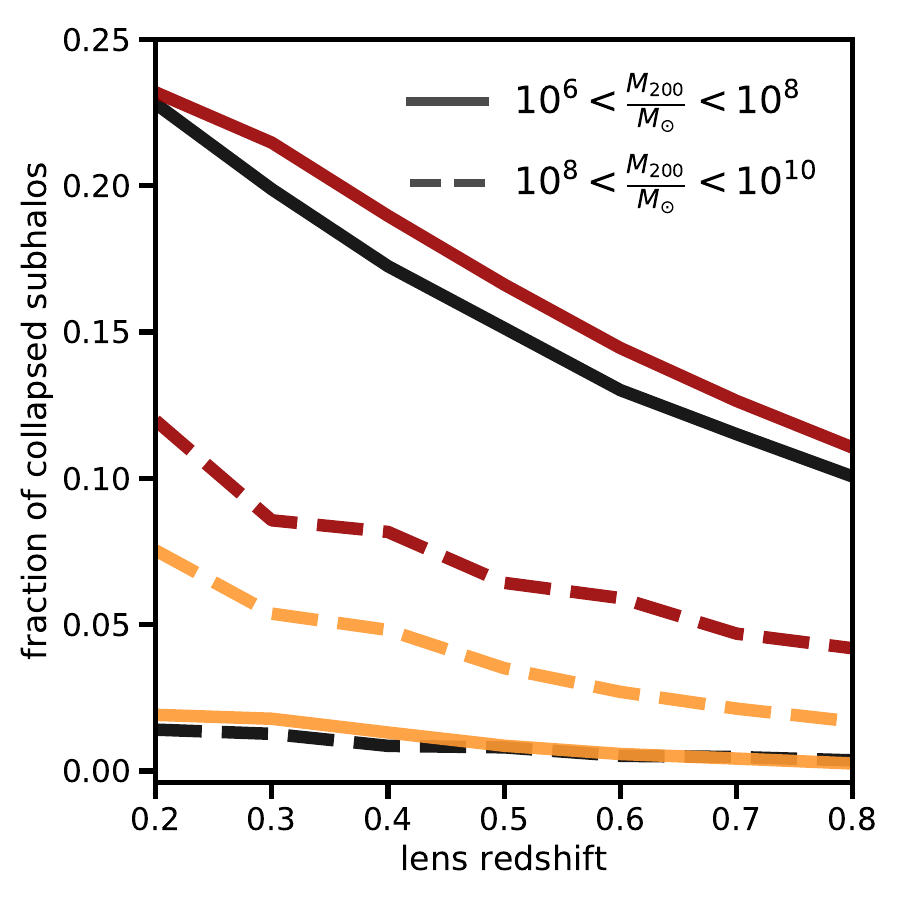}
		\caption{\label{fig:fcollapsedsubs} The fraction of core collapsed subhalos as a function of the main deflector redshift. Solid curves show the fraction of collapsed halos in the mass range $10^6 - 10^8 \msun$, while dashed curves show the fraction of collapsed subhalos in the mass range $10^8 - 10^{10}\msun$. The colors correspond to the same cross section models depicted in Figures \ref{fig:cross_section}, \ref{fig:coresizes}, and \ref{fig:tchalomass}.}
	\end{figure}
	To compute the cored density profiles of SIDM halos we follow a simple Jeans modeling approach first introduced by \citet{Kaplinghat++16}. This framework, although subject to a few internal inconsistencies \citep[e.g.][]{Sokolenko++18}, accurately reproduces the properties of SIDM halos in simulations  \citep{Robertson++20}. The relative lack of baryons inside dark matter halos on the scales relevant for strong lensing $\left(10^6 - 10^{10} \msun\right)$ somewhat simplifies the modeling, as we can safely neglect the baryonic mass component of the halos when solving for the density profile. We do, however, account for tidal forces from the baryonic matter of a central galaxy when considering core collapsed halos in Section \ref{ssec:corecollapse}.
	
	The radial Jeans equation for the density profile $\rho_{\rm{iso}}\left(r\right)$ of a fully thermalized SIDM halo with constant velocity dispersion $v_{\rm{rms}}$ is $v_{\rm{rms}}^2 \nabla \rho_{\rm{iso}} = -\rho_{\rm{iso}} \nabla \Phi$. Combining this with the Poisson equation results in a differential equation for $\rho_{\rm{iso}}$
	\begin{equation}
	\label{eqn:diffeq}
	v_{\rm{rms}}^2 \nabla^2 \ln \rho_{\rm{iso}} = -4 \pi G \rho_{\rm{iso}},
	\end{equation}
	which we may solve with boundary conditions $\rho_{\rm{iso}}^{\ \prime} \left(0\right) = 0$ and $\rho_{\rm{iso}} \left(0\right) = \rho_0$. This model has two unknowns: the central velocity dispersion $v_{\rm{rms}}$, and the central core density $\rho_0$. 
	
	To make progress, we divide the halo into two regions demarcated by a characteristic radius $r_1$. Inside $r_1$ dark matter particles scatter more frequently than once per the halo age $t_{\rm{halo}}$, while outside of this radius self-interactions do not efficiently transfer momentum between particles and the profile resembles that of a collisionless cold dark matter Navarro-Frenk-White (NFW) \citep{Navarro++96} halo with density $\rho_{\rm{NFW}}\left(r\right)$. The $r_1$ radius satisfies
	
	\begin{equation}
	\label{eqn:ratetime}
	\rho_{\rm{NFW}} \left(r_1\right) \langle \sigma v \rangle \times t_{\rm{halo}}\left(z\right) = 1,
	\end{equation}
	where $\langle \sigma v \rangle$ determines the scattering rate computed by integrating the cross section weighted by the relative velocity $v$ over a Maxwell-Boltzmann distribution \citep{Shen++21}
	\begin{equation}
	\label{eqn:thermalavg}
	\langle \sigma v \rangle = \frac{1}{2\sqrt{\pi} v_{\rm{rms}}^3}\int_0^{\infty} \sigma \left(v\right) v \times  v^2 \exp \left(\frac{-v^2}{4 v_{\rm{rms}}^2}\right) dv.
	\end{equation}
	We determine the halo age $t_{\rm{halo}}$ as a function of redshift by assuming the objects collapse at $z = 10$, a typical formation epoch for the halos in the mass range we consider, and compute the elapsed time until redshift $z$. The difference in median collapse times between halos in the mass range we consider is much less than the age of the Universe, so accounting for a more precise collapse time would introduce only a weak mass dependence in the model. We note that \citet{Robertson++20} used the NFW profile velocity dispersion in place of $v_{\rm{rms}}$ in Equation \ref{eqn:thermalavg}, while we follow the original paper by \citet{Kaplinghat++16} and use $v_{\rm{rms}}$. In practice, this choice results in only percent-level differences in the resulting core size. 
	
	In addition to Equation \ref{eqn:ratetime}, we impose two additional conditions that allow us to simultaneously solve for the unknowns $\rho_0$, $r_1$, and $v_{\rm{rms}}$. First, a solution to Equation \ref{eqn:diffeq} must have the same enclosed mass within $r_1$ as the NFW profile would have had in the absence of interactions. Second, the density profile that satisfies Equation \ref{eqn:diffeq} must match the amplitude of the NFW profile at $r_1$
	\begin{eqnarray}
	\label{eqn:massmatch}
	M_{\rm{iso}}\left(r < r_1\right) &=& M_{\rm{NFW}}\left(r < r_1\right) \\
	\label{eqn:densitymatch} \rho_{\rm{iso}}\left(r_1\right) &=& \rho_{\rm{NFW}}\left(r_1\right).
	\end{eqnarray}
	To solve these equations for a given cross section, halo mass, and redshift, we use a grid-based search method that iteratively moves through two dimensional parameter space of $\log_{10} \left(\rho_0\right)$ and $v_{\rm{rms}}$ until we match the boundary conditions at $r_1$ to $1\%$ precision\footnote{As pointed out by \citet{Kaplinghat++16} and \citet{Robertson++20}, this approach admits two solutions for the central density and velocity dispersion. We keep the solution with lower central density, reproducing the cored density profiles of halos in simulations.}. 
	
	Figure \ref{fig:densityprof} shows one solution to Equations \ref{eqn:diffeq} and \ref{eqn:ratetime} that satisfies the boundary conditions in Equations \ref{eqn:massmatch} and \ref{eqn:densitymatch}, assuming a cross section specified by $\sigma_0 = 20 \  \rm{cm^2} \ \rm{g^{-1}}$ and $v_0 = 30 \ \rm{km} \ {s^{-1}}$. The red curve shows the isothermal profile obtained from numerically solving Equation \ref{eqn:denistyprof} out to $ r = r_1$, and the gray curve shows the original NFW profile. We interpolate between the two profiles with a cored, truncated NFW profile parameterized as
	\begin{equation}
	\label{eqn:denistyprof}
	\rho \left(x, \beta, \tau \right) = \frac{\rho_s}{\left(x^{a}+ \beta^{a}\right)^{\frac{1}{a}} \left(1+x\right)^2} \frac{\tau^2}{\tau^2 + x^2},
	\end{equation}
	with $x = \frac{r}{r_s}$, truncation $\tau = \frac{r_t}{r_s}$, and core radius $\beta = \frac{r_c}{r_s}$. The parameter $a = 10$ serves to reproduce the rapid transition from a constant central density core to the NFW profile near $r_1$. Given the central core density $\rho_0$, setting $\rho\left(0, \beta, \tau\right) = \rho_0$ gives
	\begin{equation}
	\frac{r_c}{r_s} = \frac{\rho_s}{\rho_0}
	\end{equation}
	from which we obtain the core radius $r_c$ in terms of $\rho_0$ for an NFW halo with parameters $r_s$ and $\rho_s$. We determine $r_s$ and $\rho_s$ for a halo with a mass definition of 200 times the critical density at the halo redshift using the mass-concentration relation presented by \citet{DiemerJoyce19} with a scatter of 0.2 dex \citep{Wang++20} to determine the shape of the SIDM halo density profile beyond $r_1$. 
	
	Figure \ref{fig:coresizes} shows $\beta$ for the three cross section models shown in Figure \ref{fig:cross_section}. The amplitude of the cross section increases $\beta$ on all mass scales, while the velocity dependence determines the slope of the $\frac{r_c}{r_s}$ relation as a function of halo mass. For $v < v_0$, cross sections can reach amplitudes significantly larger than $10 \ \rm{cm^2}\rm{g^{-1}}$ resulting in core sizes of order half the scale radius. Very efficient scattering, generally for cross sections $> 10 \  \rm{cm^2} \ \rm{g^{-1}}$, facilitates efficient heat transfer throughout the density profile that rushes SIDM halos towards their ultimate fate. 
	
	\subsection{Core-collapsed SIDM halos}
	\label{ssec:corecollapse}
	Runaway core collapse abruptly terminates the gradual buildup of an SIDM halo's central core after heat transfer from the halo's outskirts can no longer support its growth. Many factors can impact the temporal evolution of SIDM halos, in particular the concentration of the halo and the degree to which it experiences tidal disruption by a central potential \citep{Sameie++18,Nishikawa++20,Correa21}. Dissipative self-interactions can also accelerate the onset of core collapse relative to non-dissipative interactions \citep{Essig++19,Shen++21}. 
	
	We base our model for core collapse on the results presented by \cite{Nishikawa++20}, who tracked the structural evolution of a $10^{10} \msun$ halo with and without tidal truncation in terms of a characteristic timescale for structural evolution $t_0^{-1} = \frac{4}{\sqrt{\pi}} \sigma_0 v_s \rho_s$ with $v_s = \sqrt{4 \pi G \rho_s r_s^2}$. Longer timescales correspond to slower structural evolution, and hence the time it takes for a halo to experience collapse varies proportionally with $t_0$. To apply the results presented by \citet{Nishikawa++20} to substructure lensing, we need a form for $t_0$ that accounts for the velocity dependence of the interaction cross section. To this end, we note that the timescale used by \citet{Nishikawa++20} is the same as the relaxation time $t_r^{-1} = \langle \sigma v\rangle \rho_s$ up to a constant factor of $\sim \frac{1}{3}$ \citep{Balberg++02}, where the factor of $3$ comes from the definition of $v_s$ used by \citet{Nishikawa++20} in terms of $\rho_s$ and $r_s$ instead of the velocity dispersion $v_{\rm{rms}}$. For a velocity independent cross section, computing the thermal average gives $\langle \sigma\left(v\right) v \rangle = \frac{4}{\sqrt{\pi}} \sigma_0 v_{\rm{rms}} \sim \frac{4}{3\sqrt{\pi}} \sigma_0 v_s$, so we can reasonably generalize the timescale used by \citet{Nishikawa++20} with the expression
	\begin{equation}
	\label{eqn:collapsetimescale}
	t_0 =  \frac{1}{3 \rho_s \langle \sigma v\rangle}.
	\end{equation}
	\citet{Colquhoun++20} argue that more relevant thermal averages for structure formation in SIDM models have the form $\langle \sigma\left(v\right) v^n\rangle$ with $n=2$ and $n=3$ corresponding to the rate of momentum and energy transfer, respectively. However, at the time of writing we have not found a study that examined core collapse over a wide range in halo masses using the momentum or energy exchange averages instead of the relaxation time, so we do the computation in terms of $t_r^{-1} \propto \langle \sigma\left(v\right) v \rangle$. 
	
	Figure \ref{fig:tchalomass} shows the collapse timescale $t_0$ computed with Equation \ref{eqn:collapsetimescale} as a function of halo mass for the three cross sections shown in Figure \ref{fig:cross_section}, using the same color scheme. The timescale has a strong dependence on the velocity dependence $v_0$. Halos of mass $10^{7} \msun$ have an evolution timescale roughly a factor of ten shorter than the evolution timescale in halos with mass $10^{10} \msun$. In such a model, the completely dark halos probed by substructure lensing would make up the overwhelming majority of core collapsed objects. 
	\begin{figure}
		\includegraphics[clip,trim=0cm 0cm 0cm
		0cm,width=.48\textwidth,keepaspectratio]{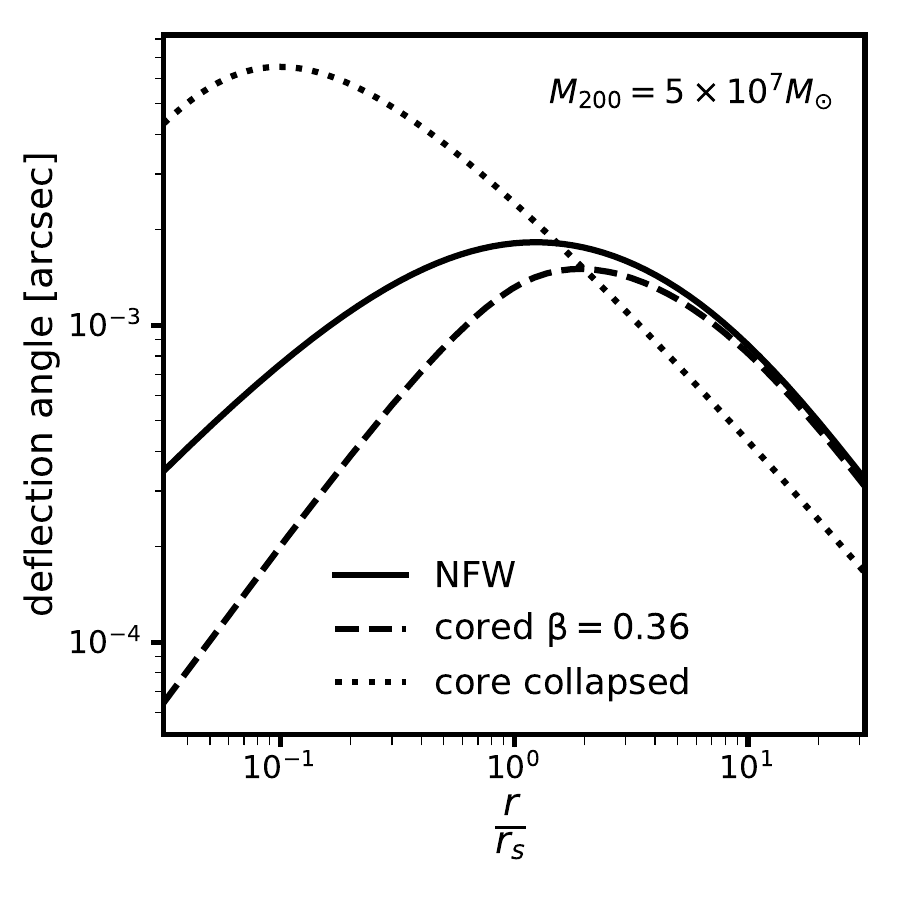}
		\caption{\label{fig:defangle} The deflection angles corresponding to the NFW, cored NFW, and core collapsed halo density profiles shown in Figure \ref{fig:densityprof}. The dashed black curve illustrates how a core with $\beta = 0.36$ suppresses the deflection at small radii relative to the NFW profile (solid black). The dotted curve shows the deflection angle corresponding to the power law density profile (Equation \ref{eqn:collapseprofile}) the halo would obtain if it underwent core collapse in our model.}
	\end{figure}	
	
	\begin{figure*}
		\includegraphics[clip,trim=0.6cm 2cm 0.5cm
		2cm,width=.96\textwidth,keepaspectratio]{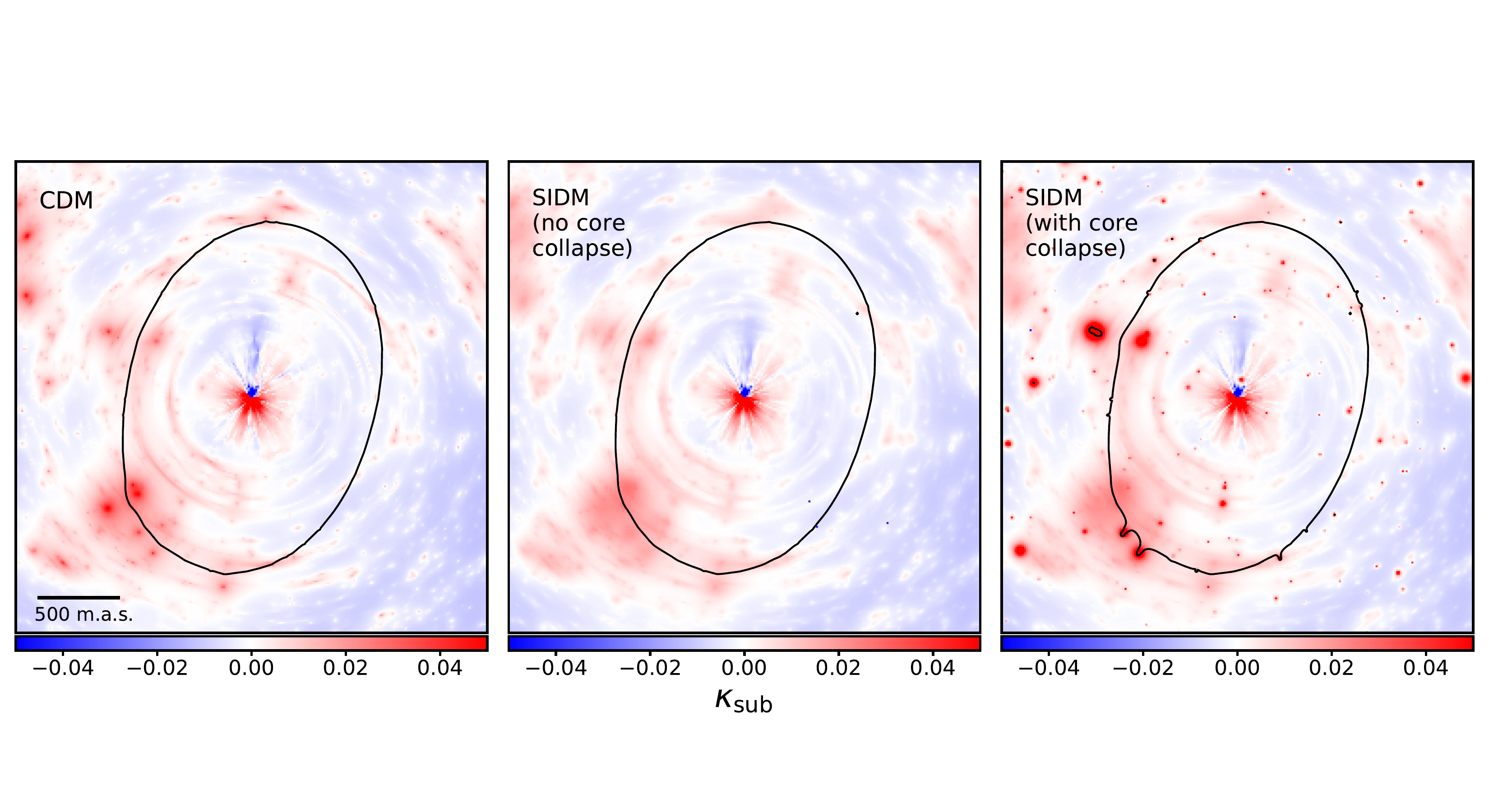}
		\caption{\label{fig:SIDMkappa} The effective multi-plane convergence in substructure $\kappa_{\rm{sub}}$, or a two dimensional representation of a lensed three dimensional mass distribution, that includes subhalos and line of sight halos. The color scale corresponds to the density in substructure relative to the average dark matter density of the Universe. Non-linear coupling between line of sight halos and the main deflector, which we parameterize as a power-law ellipsoid, manifest as projected mass arcs. Critical curves are shown as solid black lines in each panel. We place the main deflector at $z = 0.5$, and the source at $z = 1.5$. {\bf{Left:}} A realization of CDM subhalos and line of sight halos. {\bf{Center:}} The same population of CDM halos as they would appear with self-interacting dark matter core formation, but no core collapse. The cross section model used in the figure has $\sigma_0 = 35 \ \rm{cm^2} \rm{g^{-1}}$ and $v_0 = 30 \ \rm{km} \ \rm{s^{-1}}$. {\bf{Right:}} The same population of SIDM halos shown in the central panel, but now allowing some fraction of subhalos and field halos to core collapse using the model presented in Section \ref{ssec:corecollapse}. Core collapsed halos imprint clearly visible distortions in the critical curves, and occasionally produce their own critical curves and caustics.}
	\end{figure*}	
	
	\begin{figure}
		\includegraphics[clip,trim=0cm 0cm 0cm
		0cm,width=.48\textwidth,keepaspectratio]{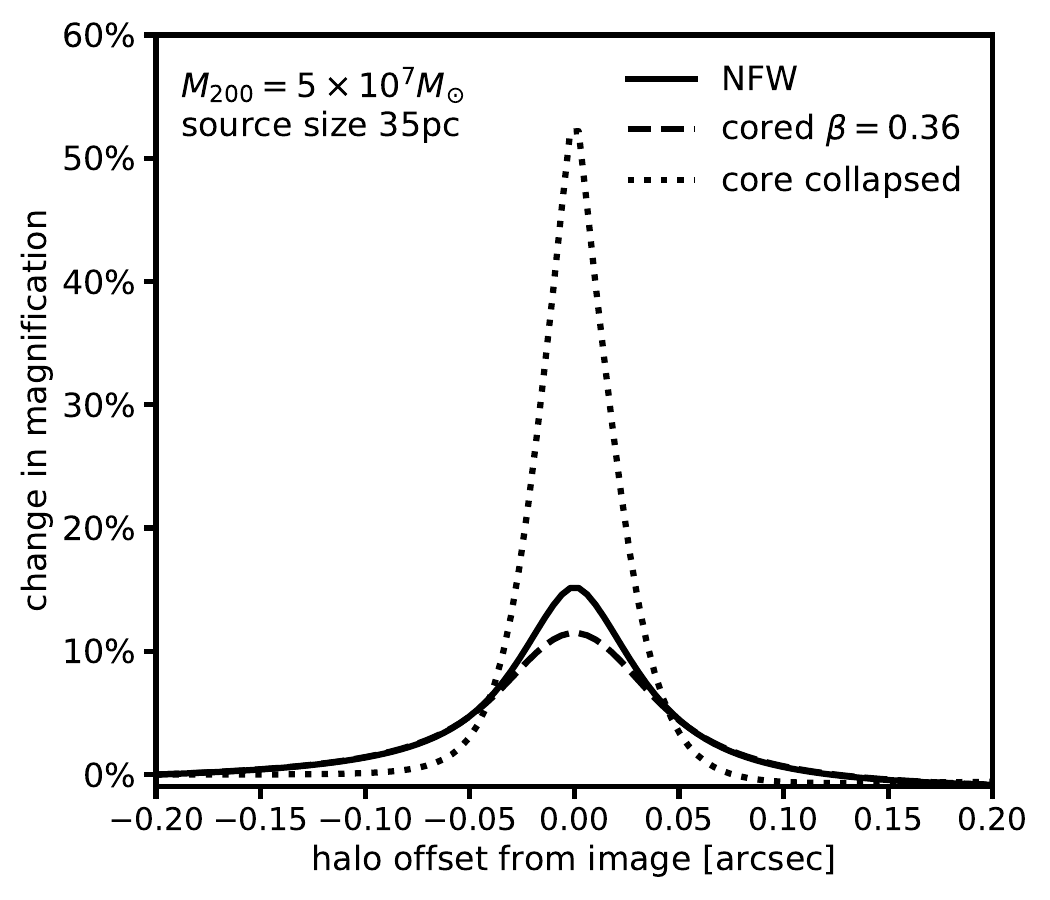}
		\caption{\label{fig:magnification}  The magnification cross sections computed by ray-tracing using the deflection angles displayed in Figure \ref{fig:defangle}. The y-axis shows the change in image magnification as a function of the angular separation of the center of the halo from a lensed image. The solid black, dashed, and dotted curves correspond to the NFW, cored NFW, and core collapsed NFW profiles, respectively. We assume a source size of 35 pc, comparable to the size of the nuclear narrow-line region of a quasar.}
	\end{figure}
	
	\citet{Nishikawa++20} find that an initial NFW halo that does not experience tidal forces during its lifetime collapses after $\mathcal{O}\left(100\right)  t_0$. In agreement with other analyses \citep[e.g.][]{Sameie++19}, \citet{Nishikawa++20} show that a tidally truncated halo core collapses more quickly, after only $\mathcal{O}\left(10\right)  t_0$. In the context of strong lensing, we associate the tidally stripped halo considered by \citet{Nishikawa++20} with a subhalo of the main deflector's host dark matter halo, while the halo that experienced no tidal stripping corresponds to a halo in the field, or in lensing terminology, a `line of sight halo'. Motivated by the results presented by \citet{Nishikawa++20}, we therefore assign a core collapse timescale to subhalos of $t_{\rm{sub}} = 10 t_0$ and field halos $t_{\rm{field}} = 100 t_0$. After computing $t_0$ for each halo in the lens system, we assign a probability of core collapse
	\begin{equation}
	\label{eqn:collapseprob}
	p_{\rm{collapse}} = \left\{
	\begin{array}{ll}
	0 & \quad t_{\rm{halo}} \leq \frac{1}{2} t^{\prime} \\
	\frac{t_{\rm{halo}} -  t^{\prime}}{ t^{\prime}} & \quad t_{\rm{halo}} \leq 2   t^{\prime} \\
	1 & \quad t_{\rm{halo}} > 2  t^{\prime},
	\end{array}
	\right.
	\end{equation}
	where $ t^{\prime}$ is either $t_{\rm{sub}}$ or $t_{\rm{field}}$, and $t_{\rm{halo}}$ is the elapsed time since the halo collapsed, the same quantity that appears in Equation \ref{eqn:ratetime}. The collapse time window around the characteristic value of $t_0$ accounts for different merger or tidal stripping histories that may accelerate or decelerate core collapse for individual halos. In principle, both $t_{\rm{sub}}$ and the distribution of collapse times around $t_{\rm{sub}}$ should depend on the structural properties of subhalos and the tidal fields in which they orbit, and future work could investigate these possibilities in more detail. 
	
	Figure \ref{fig:fcollapsedsubs} shows the fraction of core-collapsed subhalos as a function of the lens redshift predicted by our model for the same cross sections shown in Figure \ref{fig:cross_section}. The solid curves show the fraction of collapsed subhalos with mass between $10^6-10^8 M_{\odot}$, and the dashed curves show the fraction of collapsed subhalos with mass between $10^8-10^{10} M_{\odot}$; the color scheme is the same as in Figure \ref{fig:cross_section}. The fraction of collapsed subhalos at all masses decreases with the lens redshift because the halo age decreases as the lens redshift increases, and by Equation \ref{eqn:collapseprob} will have a lower probability of experiencing core collapse. Figure \ref{fig:fcollapsedsubs} suggests that a sample of strong lenses with deflectors at different redshifts could provide a unique window on the temporal evolution of SIDM density profiles, an aspect of the structure formation model that would clearly distinguish SIDM from CDM, and that strong lensing could, in principle, detect. We do not include field halos in Figure \ref{fig:fcollapsedsubs} because the model presented in this section predicts that $< 5\%$ of them collapse within a Hubble time. 
	
	The velocity dependence of the cross section determines the fraction of collapsed halos as a function of halo mass. The model shown in yellow predicts a larger fraction of collapsed subhalos in the mass range $10^8 - 10^{10} \msun$ than in the range $10^6 - 10^8\msun$, comparing the dashed yellow curve to the solid yellow curve, because $t_0$ reaches a minimum around $10^9 \msun$. This distinguishes the cross section represented in yellow from the models shown in black and red, which have more efficient scattering in low mass halos, and therefore predict a higher fraction of collapsed objects in the mass range $10^6-10^8\msun$ than in the range $10^8 - 10^{10} \msun$. 
	
	In order to compute the lensing properties of core collapsed halos, we require a model for their density profile. Core collapsed halos have central density cusps significantly steeper than the $r^{-1}$ cusps of NFW profiles \citep{Balberg++02,Sameie++19,Zavala++19,Nishikawa++20,Turner++20}. With this in mind, we model these objects using a cored power-law profile
	\begin{equation}
	\label{eqn:collapseprofile}
	\rho\left(r\right) =  \rho_c \left(1+\frac{r^2}{b^2}\right)^{-\frac{\gamma}{2}}.
	\end{equation}
	with a logarithmic profile slope $\gamma$ approximately equal to $-3$ \citep{Turner++20}. We include a small core radius $b$ to keep the total mass inside $r_s$ finite, and normalize the density profile such that the mass of the core collapsed halo inside the radius of maximum circular velocity $2.16 r_s$ matches the mass interior to $2.16 r_s$ the NFW profile would have had without self-interactions, a choice that results in halo density profiles that resemble core collapsed objects in simulations. The dotted line in Figure \ref{fig:densityprof} shows the density profile corresponding to Equation \ref{eqn:collapseprofile} that results from this procedure. 
	
	\section{Strong lensing signatures of SIDM halos}
	\label{sec:lensingbysidm}
	In this section, we compute the gravitational lensing deflection angle and magnification cross section for the SIDM density profiles described in Section \ref{sec:structuremodels}, and compare them with the lensing properties of CDM halos. We also discuss the multi-plane lensing formalism used in the forward model framework discussed in Section \ref{sec:setupandpriors}. 
	
	\subsection{Deflection angles}
	
	Projecting the three dimensional mass distribution of the halo onto the plane of the lens and then computing the projected mass in two dimensions gives the lensing deflection angle $\alpha$ for a spherical mass distribution
	\begin{equation}
	\label{eqn:deflectionintegral}
	\alpha \left(r\right) \propto \frac{1}{r} \int_0^{r} r_{\rm{2D}} dr_{\rm{2D}} \int_{-\infty}^{\infty} \rho (\sqrt{r_{\rm{2D}}^2 + z^2}) dz.
	\end{equation}
	Most spherical power law mass profiles admit analytic solutions for $\alpha$, but we could not find a closed form solution to the integrals for the cored density profile in Equation \ref{eqn:denistyprof}. To perform lensing computations with the cored profile, we integrate Equation \ref{eqn:deflectionintegral} numerically for values of $\beta$ and $\tau$ between $\left(0.01, 1\right)$ and $\left(1, 35\right)$, respectively, up to a constant numerical factor. We then interpolate the solutions to obtain the shape of the deflection angle curve as a function of $\frac{r}{r_s}$, and normalize such that the deflection angle of the cored profile at $30 r_s$ equals that of a truncated NFW profile \citep{Baltz++09} when $\beta = 0.01$. 
	
	In Figure \ref{fig:defangle}, we plot the deflection angle produced by a $5 \times 10^{7} \msun$ halo at $z=0.5$, using the same line styles to identify the profiles as in Figure \ref{fig:densityprof}. The presence of a core suppresses the deflection angle at small radii, while the deflection angle produced by the core-collapsed object is roughly an order of magnitude larger the the original NFW halo for $r < r_s$. 
	
	\begin{table*}
		\centering
		\caption{Parameters sampled in the forward model, and quantities derived from them. $\mathcal{N}$ indicates a Gaussian prior, and $\mathcal{U}$ indicates a uniform prior.}
		\label{tab:params}
		\begin{tabular}{lccr} 
			\hline
			parameter & description & prior\\
			\hline 
			$\sigma_0 \left[\rm{cm^2} \ \rm{g^{-1}}\right]$ & asymptotic value of the interaction cross &  $\mathcal{U}$ $\left(0.5, 50\right) $\\
			& section at low velocity (Equation \ref{eqn:crossec}) & \\
			\\
			$\sigma_{20} \left[\rm{cm^2} \ \rm{g^{-1}}\right]$ & cross section amplitude at $20 \ \rm{km} \ \rm{s^{-1}}$ &  (derived quantity) \\
			\\
			$v_0 \left[\rm{km}\  \rm{s^{-1}}\right]$ & velocity scale of the  SIDM cross section &  $\mathcal{U}$ $\left(10, 50\right) $\\
			& $\sigma\left(v\right) \propto v^{-4}$ for $v > v_0$ (Equation \ref{eqn:crossec}) & \\
			\\
			$b$ & core size in units of $r_s$ of core collapsed halos &  $\mathcal{U}$ $\left(0.01, 0.05\right) $\\&(Equation \ref{eqn:collapseprofile})& \\
			\\
			$\gamma$ & logarithmic slope of core collapsed &  $\mathcal{U}$ $\left(2.9, 3.1\right) $\\& halo density profiles (Equation \ref{eqn:collapseprofile})& \\
			\\
			$\Sigma_{\rm{sub}} \left[\rm{kpc}^{-2}\right]$ &subhalo mass function normalization (Equation \ref{eqn:subhalomfunc})&  \\&tidal stripping efficiency $0.5 \times$ Milky Way& $\mathcal{N}$ $\left(0.050, 0.010\right) $\\
			& tidal stripping efficiency $0.75 \times$  Milky Way & $\mathcal{N}$ $\left(0.032, 0.007\right) $\\
			\\
			$\alpha$ & logarithmic slope of subhalo mass function&  $\mathcal{U}$ $\left(-1.95, -1.85\right) $\\&(Equation \ref{eqn:subhalomfunc})& \\
			\\
			$\delta_{\rm{los}}$ & rescales the line of sight halo mass function & $\mathcal{U}$ $\left(0.8, 1.2\right) $\\
			&$10^6 < m < 10^{10} \msun$ (Equation \ref{eqn:losmfunc})&\\
			\\
			$\sigma_{\rm{src}} \left[\rm{pc}\right]$ & background source size &  \\ &nuclear narrow-line emission&$\mathcal{U}\left(25, 60\right)$\\ 
			& mid-IR emission & $\mathcal{U}\left(0.5, 5\right)$ \\&
			\\
			$\gamma_{\rm{macro}}$ & logarithmic slope of main deflector mass profile  & $\mathcal{U}$ $\left(1.9, 2.2\right) $\\
			\\
			$a_4$ & controls boxyness/diskyness of main& $\mathcal{N}$ $\left(0, 0.01\right) $ \\&deflector mass profile & \\
			\\
			$\delta_{xy} \left[\rm{m.a.s.}\right]$ & image position measurement uncertainty& $\mathcal{N}$ $\left(0, 3\right) $\\
			\\
			$\delta f $ & image flux measurement uncertainties& \\
			& mid-IR & $2\%$ \\
			& narrow-line & $4\%$ \\
			\hline		
			
		\end{tabular}
	\end{table*}
	
	Real lens systems include many subhalos and line of sight halos. To compute observables in this scenario we use the multi-plane lens equation \citep{BlandfordNarayan86}
	\begin{eqnarray}
	\label{eqn:raytracing}
	\boldsymbol{\theta_K}\left(\boldsymbol{\theta}\right) &=& \boldsymbol{\theta} - \frac{1}{D_{\rm{s}}} \sum_{k=1}^{K-1} D_{\rm{ks}}{\boldsymbol{\alpha_{\rm{k}}}} \left(D_{\rm{k}} \boldsymbol{\theta_{\rm{k}}}\right) \\
	\nonumber & \equiv & \boldsymbol{\theta} - \boldsymbol{\alpha}_{\rm{eff}}\left(\boldsymbol{\theta}\right)
	\end{eqnarray} 
	where $\boldsymbol{\theta_K}$ is the angular coordinate of a light ray on the $K$th lens plane, $\boldsymbol{\theta}$ is a coordinate on the sky, $D_{\rm{k}}$ is the angular diameter distance to the $k$th lens plane, $D_{\rm{ks}}$ is the angular diameter distance from the $k$th lens plane to the source plane, and $\boldsymbol{\alpha_k}$ is the deflection field at the $k$th lens plane. We have also defined the effective deflection angle $\boldsymbol{\alpha}_{\rm{eff}} \equiv \frac{1}{D_{\rm{s}}} \sum D_{\rm{ks}}{\boldsymbol{\alpha_{\rm{k}}}} \left(D_{\rm{k}} \boldsymbol{\theta_{\rm{k}}}\right)$. 
	
	Figure \ref{fig:SIDMkappa} shows the effective multi-plane convergence, a two dimensional representation of a lensed three dimensional mass distribution, for a single realization of subhalos and line of sight halos. We define the effective multi-plane convergence from substructure as $\kappa_{\rm{sub}} \equiv \frac{1}{2}\div \boldsymbol{\alpha}_{\rm{eff}} - \kappa_{\rm{macro}}$, where $\kappa_{\rm{macro}}$ is the convergence from a main lens profile that we have parameterized as a power-law ellipsoid. The left panel shows a population of CDM halos modeled as NFW profiles. The central panel shows how the same population of halos would appear if dark matter had an interaction cross section given by Equation \ref{eqn:crossec} with $\sigma_0 = 35 \ \rm{cm^2} \ \rm{g^{-1}}$ and $v_0 = 30 \  \rm{km} \ \rm{s^{-1}}$; core collapse is `turned off' in the central panel, effectively setting $t_0 = \infty$ for every halo. The far right panel shows the same realization as the central panel, but now allowing a fraction of halos to core collapse through the modeling approach described in Section \ref{ssec:corecollapse}. The pronounced distortions in the critical curve (shown in black) clearly distinguish core collapsed objects from cored and cuspy NFW profiles. Some core collapsed subhalos even have critical curves completely detached from the main deflector's caustic network, indicating that they have super-critical central densities and can produce multiple images. 
	
	\subsection{Magnification cross section}
	Investigations of substructure in multiply imaged quasar strong lens systems rely on the magnification ratios\footnote{Magnification ratios factor out the unknown source brightness, and only depend on the relative distortions between image pairs.} (or flux ratios) between images to reveal the presence of otherwise invisible dark matter structure \citep{MaoSchneider98,Metcalf++01,D+K02,Nierenberg++14,Nierenberg++17,Hsueh++20,Gilman++20a,Gilman++20b}. The utility of these data stems from their dependence on the second derivatives of the projected gravitational potential, resulting in an extremely sensitive localized probe of substructure. 
	
	Figure \ref{fig:magnification} shows the the magnification cross section for a subhalo with mass $5 \times 10^{7} \msun$ modeled as an NFW profile (solid), a cored NFW profile (dashed), and the profile the subhalo would have after collapse using the model presented in Section \ref{ssec:corecollapse}. A core radius of $0.36 r_s$ reduces the peak magnification perturbation by $\sim 20\%$ relative to an NFW profile while leaving the magnification cross section beyond 0.05 arcseconds unchanged. The core collapsed halo changes the peak magnification by over $300 \%$ when the halo lies on top of an image in projection, but creates a weaker perturbation than the NFW profile at larger projected distances. In Appendix \ref{sec:appA}, we show how the magnification cross section depends on the logarithmic slope of the core collapsed profile, and on the radius at which we match the enclosed mass between it and the NFW profile the same halo would have in CDM.
	
	The significant difference between the magnification cross sections shown in Figure \ref{fig:magnification} demonstrates that image flux ratios can, in principle, provide a means of probing the internal structure of self-interacting dark matter halos. If a core collapsed subhalo happens to align with a lensed image, given the magnitude of the perturbation from the core-collapsed halo relative to that of the NFW halo in Figure \ref{fig:magnification}, we might expect to find some lenses with flux ratios that CDM, let alone smooth lens models, cannot explain. If no direct hit by a core collapsed subhalo occurs, a population of cored halos should produce less perturbation to image magnifications relative to CDM, on average. To assess quantitatively how a sample of quadruple image lenses can disentangle these competing effects and constrain the cross section, we use a forward modeling inference pipeline to analyze mock lenses with full populations of SIDM structure included in the lens model.  
	
	\section{Inference on mock SIDM datasets: setup, methodology and modeling assumptions}
	\label{sec:setupandpriors}
	
	Using a hierarchical Bayesian inference pipeline developed by \citet{Gilman++18,Gilman++19}, we perform a joint inference on the parameters describing the SIDM cross section and the form of the (sub)halo mass function. Performing the full forward modeling inference on mock datasets enables quantitative forecasts for constraints on the cross section, reveals covariance between different model parameters, and builds physical intuition for how best to deploy lensing as a probe of SIDM on sub-galactic scales. We begin in Section \ref{ssec:inference} with a brief review of the forward modeling inference pipeline. Section \ref{ssec:models} describes the parameterizations assumed for the halo mass function, the density profiles of SIDM halos, and other model ingredients such as the lens macromodel and properties of the lensed background source. 
	
	\subsection{Inference methodology}
	\label{ssec:inference}
	The method we use to infer the parameters describing the SIDM cross section was developed and tested by \citet{Gilman++18,Gilman++19}, and was used to constrain the free-streaming length of dark matter \citep{Gilman++20a}, and the concentration-mass relation of CDM halos \citep{Gilman++20b}. We forward model flux ratios for images at the observed image coordinate by ray tracing (with Equation \ref{eqn:raytracing}) through full populations of subhalos and line of sight halos, and estimate the likelihood function of the model parameters given the observed data with summary statistics and an implementation of Approximate Bayesian Computing. We simultaneously sample hyper-parameters describing dark matter properties and nuisance parameters such as the size of the background source and parameters that describe the main deflector's mass profile, properly marginalizing over the many nuisance parameters in lensing. For the simulations presented in this work, we processed approximately 500,000 realizations per lens. We refer to \citet{Gilman++20a} for a detailed overview of the inference framework. 
	
	\subsection{Model Ingredients}
	\label{ssec:models}
	Our simulations combine models of the subhalo and field halo mass functions with the prescription for modeling SIDM halo density profiles described in Section \ref{sec:structuremodels}.  In addition, we also require models for the main deflector mass profile and the lensed background source in order to compute lensing observables. In the following subsections, we describe the models implemented for each of these components. Table \ref{tab:params} lists the parameter names, gives a brief description of them, and states the priors assumed in our simulations. 
	
	\subsubsection{The self-interaction cross section and halo density profiles}
	\label{ssec:SIDMcross}
	We use the parameterization for the SIDM cross section given in Equation \ref{eqn:crossec} with a uniform prior on $\sigma_0$ between $0.5 - 50 \ \rm{cm^2}\rm{g^{-1}}$, and a uniform prior on $v_0$ between $10 - 50 \ \rm{km} \ \rm{s^{-1}}$. This parameter space includes cross sections on the scale of dwarf galaxies that range from $50 \ \rm{cm^{2}} \rm{g^{-1}}$ when $v_0 = 50 \ \rm{km} \ \rm{s^{-1}}$, to less than $0.1 \ \rm{cm^2} \rm{g^{-1}}$ for $v_0 = 10 \ \rm{km} \ \rm{s^{-1}}$. While the combination of $\sigma_0$ and $v_0$ that gives the smallest cross section has a non-zero cross section and is therefore inconsistent with CDM, the core sizes become much smaller than $r_s$ and the fraction of core collapsed halos approaches zero, resulting in halo populations indistinguishable from CDM. 
	
	We compute the core radii for halos using the method discussed in Section \ref{ssec:coredhalos}, and allow some fraction of subhalos and field halos to core collapse using the approach outlined in Section \ref{ssec:corecollapse}. We model cored halos using the cored NFW profile in Equation \ref{eqn:denistyprof}, and core collapse halos using the cored power law profile in Equation \ref{eqn:collapseprofile}. We assume core collapsed halos have logarithmic profile slopes $\gamma$ close to $-3$, setting a uniform prior on $\gamma$ between $-2.8$ and $-3.2$ \citep{Turner++20}. We include a small core radius $b$ in core collapsed halos, using a uniform prior on $b$ between $0.01-0.05 r_s$. Current simulations lack the resolution necessary to resolve the centers of core collapse objects, but the density profile must eventually flatten in the center of the halo or the objects with $\gamma \leq-3$ would have infinite mass interior to $r_s$. We treat both $b$ and $\gamma$ as population-level parameters, assigning the same values to each halo in a particular realization.
	
	\subsubsection{The subhalo and field halo mass functions}
	We use the same parameterization for the subhalo mass function as presented by \citet{Gilman++20a}, modeled as a power-law in infall mass $m$ with logarithmic slope $\alpha$, defined in projection: 
	\begin{equation}
	\label{eqn:subhalomfunc}
	\frac{d^2 N_{\rm{sub}}}{dm dA} =  \frac{\Sigma_{\rm{sub}}}{m_0} \left(\frac{m}{m_0}\right)^{\alpha} \mathcal{F} \left(M_{\rm{halo}}, z\right),
	\end{equation}
	where 
	\begin{equation}
	\label{eqn:scaling}
	\log_{10} \left(\mathcal{F}\right) = k_1 \log_{10} \left(\frac{M_{\rm{halo}}}{10^{13} \msun}\right) + k_2 \log_{10}\left(z+0.5\right)
	\end{equation}
	is a scaling function that captures the evolution of the projected mass in substructure as a function of host halo mass $M_{\rm{host}}$ and redshift, with $k_1 = 0.88$ and $k_2 = 1.7$ \citep{Gilman++20a}. For simplicity, throughout the simulations we assume a halo mass $M_{\rm{halo}} = 10^{13.3}M_{\odot}$, the population mean halo mass for strong lenses inferred by \citep{Lagattuta++10}. 
	
	We use two different priors on the amplitude of the subhalo mass function $\Sigma_{\rm{sub}}$ that bracket a plausible range of subhalo abundance in the host dark matter halos of lensing galaxies. We define these priors in terms of the differential tidal stripping efficiency between the Milky Way and massive elliptical galaxies \citep{Nadler++21}. First, we consider $\Sigma_{\rm{sub}} = 0.05 \pm 0.01 \ \rm{kpc^{-2}}$, which corresponds to the assumption that the Milky Way disrupts halos twice as efficiently as the elliptical galaxies that typically act as strong lenses. Second, we consider a prior $\Sigma_{\rm{sub}} = 0.032 \pm 0.007 \ \rm{kpc^{-2}}$, which assumes disruption in the Milky Way is only $25 \%$ more efficient. Both of these priors are motivated by, and consistent with, existing measurements of the subhalo mass function from lensing \citep{Gilman++20a} and the Milky Way \citep{Nadler++21,Banik++21}. We use the semi-analytic modeling tool {\tt{galacticus}} \citep{Benson12} to connect the two regimes. 
	
	We model line of sight halos with
	\begin{equation}
	\label{eqn:losmfunc}
	\frac{d^2N}{dm  dV} =  \dlos \big(1+ \xi_{\rm{2halo}}\left(M_{\rm{halo}}, z\right)\big) \frac{d^2N_{\rm{ST}}}{dm  dV}.
	\end{equation}
	where $\frac{d^2N_{\rm{ST}}}{dm  dV}$ is the familiar Sheth-Tormen mass function. We allow a $20 \%$ flexibility in the amplitude of the halo mass function through $\delta_{\rm{LOS}}$ with a uniform prior between $0.8 - 1.2$ to account for possible systematics associated with the assumed model for the halo mass function, and the effects of baryons on small-scale structure formation \citep{Benson20}.  The two halo term $\xi_{\rm{2halo}}$ accounts for correlated structure around the host halo of the main deflector \citep{Gilman++19,Lazar++21}. Without the inclusion of $\xi_{\rm{2halo}}$, correlated structure would be absorbed into the inference on $\Sigma_{\rm{sub}}$. 
	
	Some self-interacting dark matter models predict both scattering between particles and a suppression in the matter power spectrum \citep[e.g.][]{Vogelsberger++16}, although the suppression of small-scale power is not a generic feature of all SIDM models. Recently, \citet{Nadler++20} showed that velocity-dependent cross sections that do not directly affect the matter power spectrum can still suppress the amplitude of the subhalo mass function if the interaction cross section has an appreciable amplitude on the velocity scale corresponding to the typical infall speed of an accreted halo $\sim 200 \ \rm{km} \ \rm{s^{-1}}$. We do not explicitly include these effects in our models for the subhalo and halo mass functions because the velocity-dependence of the cross sections we consider strongly suppresses interactions at the characteristic velocity scales of infalling halos. The reduced central densities of cored SIDM halos make them more prone to tidal stripping than cuspy CDM halos \citep{Penarrubia++10,Dooley++16}, with the effect of coupling the interaction cross section to the projected number density of subhalos. In this work, we assume the field halo mass function is the same as predicted by CDM, and absorb differential tidal stripping effects in subhalos between SIDM and CDM into the amplitude of the subhalo mass function. The lensing signal we constrain therefore derives entirely from the density profiles of halos, not their abundance relative to CDM. 
	
	\subsubsection{The main deflector lens model}
	\label{ssec:macromodel}
	We model the main deflector as a power-law ellipsoid with logarithmic profile slope $\gamma$ \citep{Tessore++15}, plus external shear. The power-law mass profile is a physically motivated mass model for the massive elliptical galaxies that typically act as strong lenses \citep{Gavazzi++07,Auger++10,Gilman++17}. We assume a uniform prior on $\gamma$ between $1.9 - 2.2$ \citep{Auger++10}. 
	
	We allow for additional flexibility in the main deflector mass profile by adding boxy and disky isodensity contours through an octopole mass distribution with projected mass
	\begin{equation}
	\label{eqn:boxyeqn}
	\kappa\left(r\right) = \frac{a_4}{r} \cos \left(4 \left(\phi - \phi_m \right)\right),
	\end{equation}
	with the orientation $\phi_m$ fixed to align with the position angle of the elliptical mass profile. The parameter $a_4$ determines the amplitude of the additional mass component, and results in boxy (disky) isodensity contours when $a_4$ is less than (greater than) zero. We assume a Gaussian prior on $a_4$ of $\mathcal{N}\left(0, 0.01\right)$ based on observations of the surface brightness contours of elliptical galaxies \citep{Bender++89,Saglia++93}. Marginalizing over the range of $a_4$ values measured from the light very likely overestimates the range of boxyness and diskyness of the total mass profile after accounting for the projected mass from the host dark matter halo. In this regard, the prior we assume for $a_4$ maximizes the flexibility of the main deflector mass profile and suppresses the signal from SIDM structure in the lens model. 
	
	\subsubsection{The background source}
	\label{ssec:backgroundsrc}
	For each realization in the forward model we compute flux ratios assuming two different sources sizes. The two source models we consider have emission from a spatially extended region in the source plane, washing out contaminating effects of microlensing by stars in the foreground galaxy. Because the perturbation by a halo of fixed mass to lensed image depends on the size of the background source \citep{DoblerKeeton06}, we must explicitly account for the finite-size background source in the forward model. 
	
	First, we consider a sample of lenses with measured narrow-line flux ratios \citep{MoustakasMetcalf02,Nierenberg++20}. Emission from the nuclear narrow-line region extends out to $\mathcal{O}\left(10\right) \rm{pc}$ \citep{MullerSanchez++11} from the central engine of the background quasar, washing out contaminating effects of microlensing by stars in the foreground galaxy. Second, we consider image fluxes measured from rest-frame mid-IR emission. Like the narrow-line emission, the spatial extent of the mid-IR emitting region renders these data immune to microlensing \citep{Sluse++13}\footnote{Microlensing can still contaminate mid-IR emission at wavelengths below 10 microns, but we consider observations at 20 microns with JWST for which microlensing leads to uncertainties of only a few percent (Sluse et al., in prep).}, but the more compact size of $\mathcal{O}\left(1\right) \rm{pc}$ makes mid-IR dataset a more sensitive probe of substructure. For both the simulated narrow-line and mid-IR datasets, we model the background source as a circular Gaussian with a prior on its full width at half maximum between $25-60 \rm{pc}$ for the narrow-line data, and between $0.5 -5 \rm{pc}$ for the mid-IR data. We marginalize over the range of source sizes specified by the prior when computing constraints from each dataset. 
	
	\subsubsection{Measurement uncertainties}
	We propagate flux and astrometric uncertainties directly through the forward model. After computing the flux ratios for each realization, we add statistical measurement uncertainties to the image fluxes $4\%$ to the narrow-line data and $2 \%$ to the mid-IR data. We chose the value of $2 \%$ for the mid-IR fluxes based on the expected precision of future measurements with JWST through observing proposal JWST GO-02046 (PI Nierenberg). The uncertainty of $4\%$ assumed for the narrow-line fluxes is the median precision of the current sample of narrow-line lenses \citep{Nierenberg++14,Nierenberg++17,Nierenberg++20}. We assume astrometric uncertainties of $3$ m.a.s., sampling perturbations to the image positions for each realization in the forward model before ray-tracing. 
	
	\subsection{Mock datasets}
	\label{ssec:mockdata}
	We generate 30 mock lenses with Einstein radii, deflector and source redshifts, external shears, and logarithmic profile slopes consistent with those expected for a sample of strong lenses \citet{OguriMarshall10,Auger++10} that we have recently obtained with HST-GO-15320 and HST-GO-15632 (PI Treu). We will obtain a sample of 30 lenses with mid-IR flux measurements in the near future through JWST GO-02046 (PI Nierenberg). Each mock has boxy or disky isodensity contours with an amplitude $a_4$ drawn from a prior with mean zero and variance 0.01. The mocks have an approximately equal number of cross, cusp, and fold image configurations. 
	
	We consider two dark matter models. First, we generate a CDM-like dataset with a cross section normalization $\sigma_0 = 0.5 \ \rm{cm^2} \rm{g^{-1}}$ and $v_0 = 10 \ \rm{km} \ \rm{s^{-1}}$. Although the model has a non-zero cross section, a small value of both $\sigma_0$ and $v_0$ lead to $t_0 >> t_{\rm{halo}}$ and $r_c << r_s$, and thus the model predicts halo populations with structural properties indistinguishable from CDM. Second, we consider an SIDM model with a normalization $\sigma_0 = 40 \ \rm{cm^2} \rm{g^{-1}}$ and $v_0 = 30 \ \rm{km} \ \rm{s^{-1}}$. The SIDM model has a cross section of $10 \ \rm{cm^2} \rm{g^{-1}}$ at $30 \ \rm{km} \ \rm{s^{-1}}$, which is consistent with current constraints on the cross section and may explain some features of dwarf galaxy stellar dynamics observed around the Milky Way \citep{Kahlhoefer++19}. 
	
	\section{Constraints on the SIDM cross section with mock strong lens datasets}
	\label{sec:results}
	We combine the inference framework described in the previous section with the model for structure formation in SIDM described in Section \ref{sec:structuremodels} to forecast constraints on the interaction cross section with 30 mock datasets. In Section \ref{ssec:constraintscdm}, we consider constraints on the cross section assuming CDM, and present forecasts for ruling out SIDM models as a function of the number of lenses and the type of flux ratio measurement (narrow-line or mid-IR). In Section \ref{ssec:constraintssidm}, we discuss the prospects for ruling out CDM, assuming SIDM. 
	\begin{figure}
		\includegraphics[clip,trim=0cm 0cm 0cm
		0cm,width=.48\textwidth,keepaspectratio]{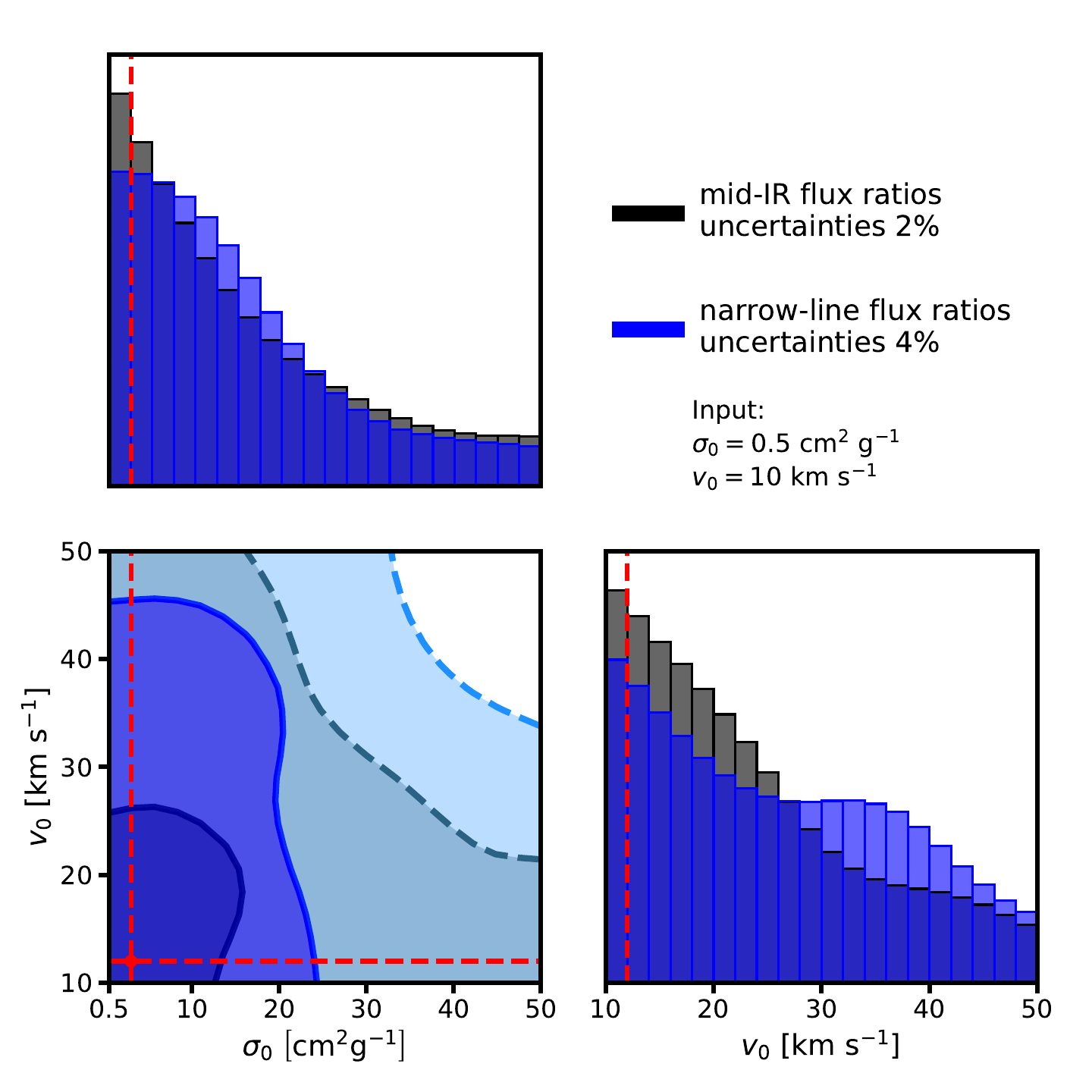}
		\caption{\label{fig:exampleconstraintcdm} Joint constraint on $\sigma_0$ and $v_0$ from 30 mock lenses assuming a CDM-like dataset with input values of $\sigma_0 = 0.5 \ \rm{cm^2} \rm{g^{-1}}$ and $v_0 = 10 \ \rm{km} \ \rm{s^{-1}}$ identified in the subplots in red. The input values of $\sigma_0$ and $v_0$ predict structural properties in halos indistinguishable from CDM with lensing. The black (blue) contours show the inference from the same set of 30 lenses with mid-IR flux ratios (narrow-line flux ratios) marginalized over the source size $0.5 - 5 \rm{pc}$ $\left(25-60 \rm{pc}\right)$ and assuming a prior on the amplitude of the subhalo mass function $\mathcal{N}\left(0.05, 0.01\right)$.}
	\end{figure}
	\begin{figure}
		\includegraphics[clip,trim=0cm 0cm 0cm
		0cm,width=.45\textwidth,keepaspectratio]{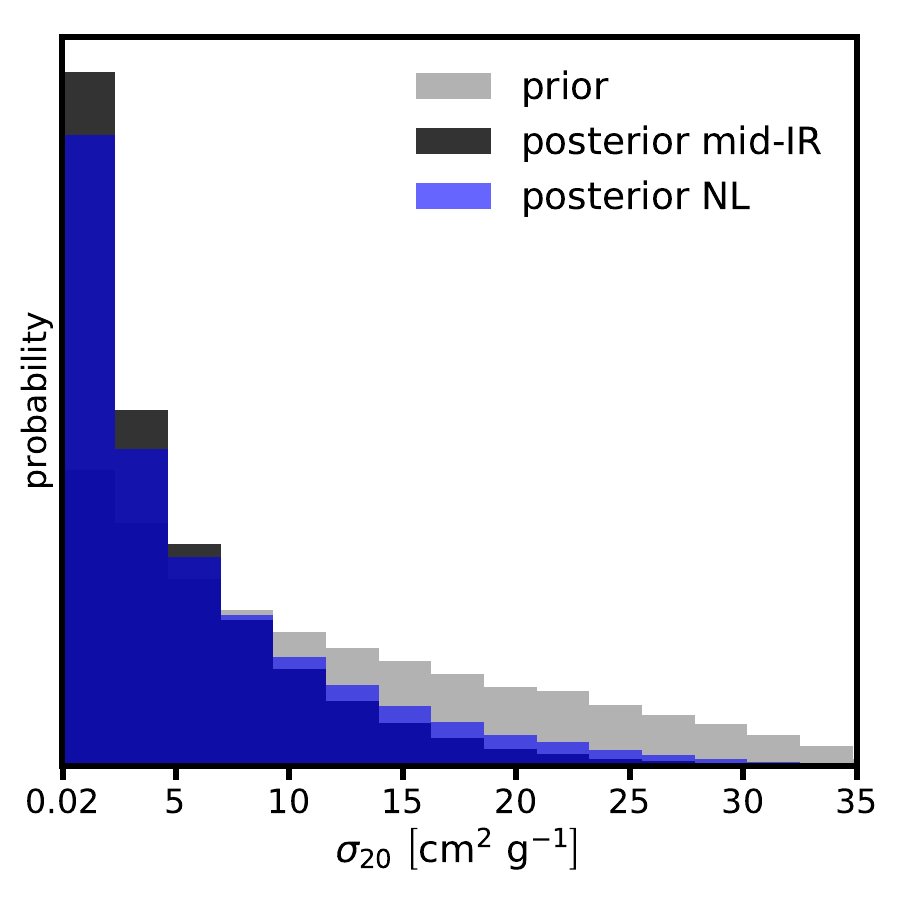}
		\includegraphics[clip,trim=0cm 0cm 0cm
		0cm,width=.45\textwidth,keepaspectratio]{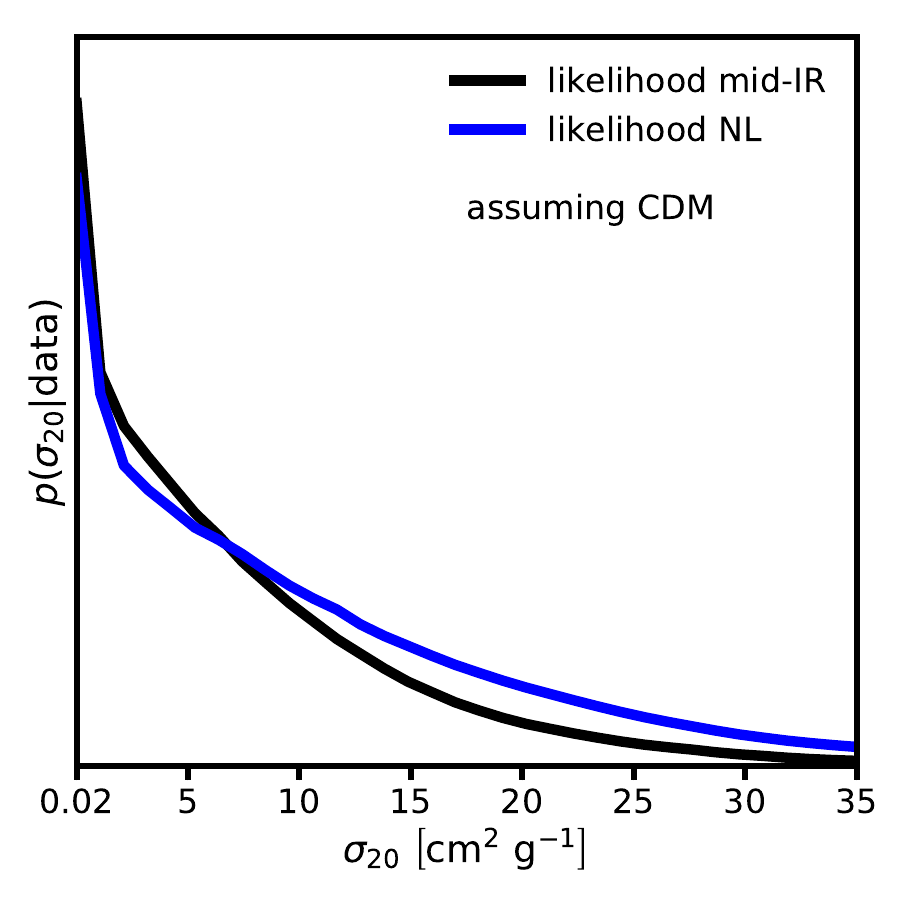}
		\caption{\label{fig:priorposteriorlikelihoodcdm} {\bf{Top:}} The gray histogram shows the effective prior on the cross section evaluated at $20 \ \rm{km} \ \rm{s^{-1}}$ $\left(\sigma_{20}\right)$ that corresponds to a uniform prior on $\sigma_0$ and $v_0$. The blue and black histograms show the posterior probability distribution of $\sigma_{20}$ inferred from the mock lenses computed by sampling the likelihood in Figure \ref{fig:exampleconstraintcdm}. {\bf{Bottom:}} The likelihood of $\sigma_{20}$ from the lensing analysis, which we compute by dividing the posterior distributions shown in the top panel by the effective prior. The curves have the interpretation of posterior distributions $p\left(\sigma_{20} | \rm{data}\right)$ assuming a uniform prior on $\sigma_{20}$.}
	\end{figure}
	\begin{figure}
		\includegraphics[clip,trim=0cm 0cm 0cm
		0cm,width=.48\textwidth,keepaspectratio]{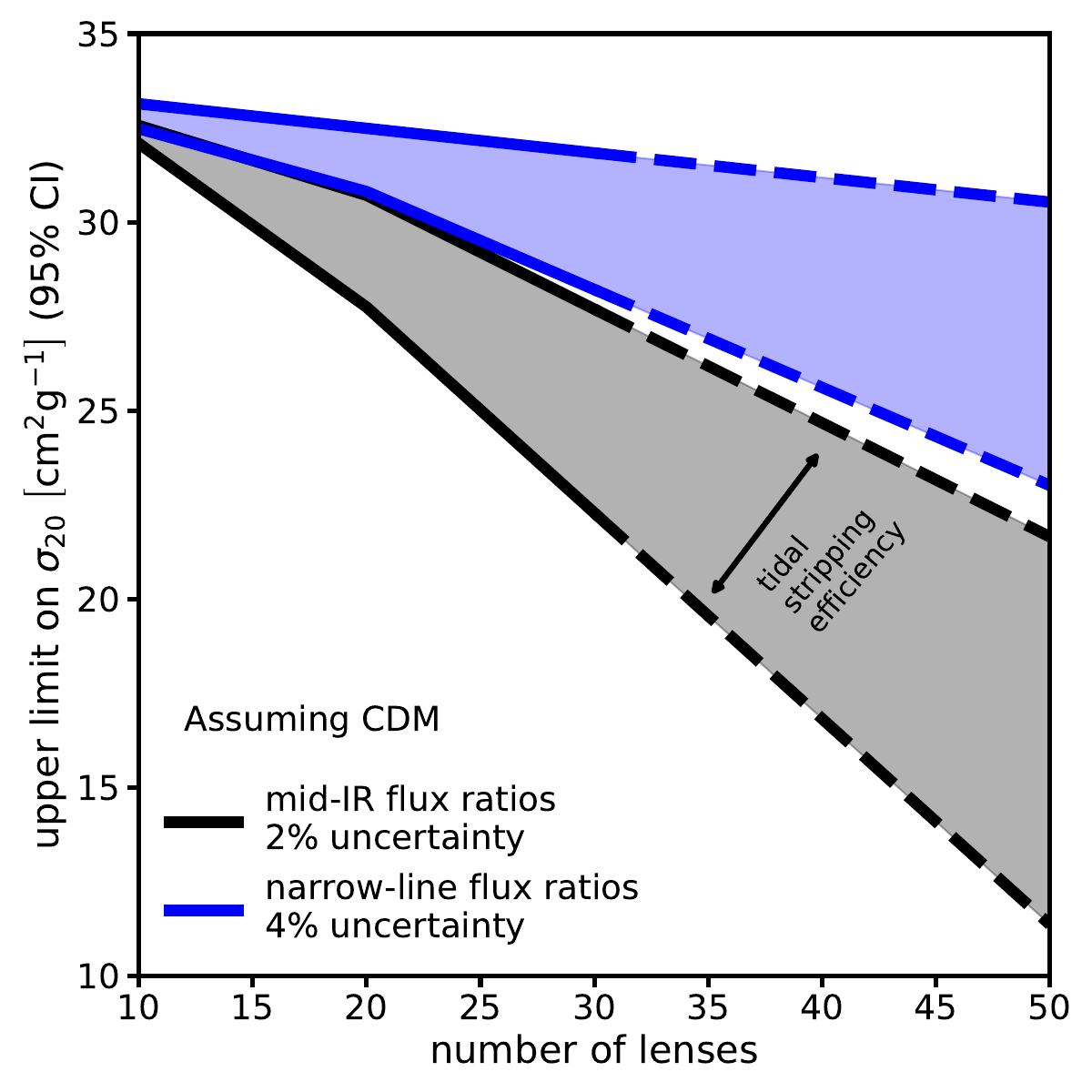}
		\caption{\label{fig:forecast} Forecasts for the constraint on $\sigma_{20}$ as a function of the number of lenses and the type of flux ratio measurement. The lower and upper ranges of the forecast correspond to priors on the subhalo mass function that assume halos are disrupted twice as efficiently in the Milky Way than in elliptical galaxies, or $\mathcal{N}\left(0.05, 0.01\right)$, and one quarter as efficiently, or $\mathcal{N}\left(0.032, 0.007\right)$, respectively \citep{Nadler++21}. The uncertainties of the flux ratio measurement indicate the expected precision for mid-IR flux measurements with JWST through proposal JWST GO-02046 (PI Nierenberg), and the median uncertainty of the current sample of narrow-line lenses.}
	\end{figure}
	\subsection{Constraints on the cross section assuming CDM}
	\label{ssec:constraintscdm}
	
	Figure \ref{fig:exampleconstraintcdm} shows a joint inference with 30 mock lenses on $\sigma_0$ and $v_0$ assuming a CDM-like model with input values of $\sigma_0$ and $v_0$ that predict halo structural properties indistinguishable from CDM, and assuming a prior on the subhalo mass function amplitude $\mathcal{N}\left(0.05, 0.01\right)$. The black contours show the $68\%$ and $95\%$ confidence intervals for a sample of 30 lenses with narrow-line fluxes measured to $4\%$ precision, while the blue contours correspond to a sample of 30 lenses with mid-IR fluxes measured to $2\%$ precision. The improved constraining power of the mid-IR dataset relative to the narrow-line dataset stems from two sources: First, JWST will measure mid-IR data more precisely than current technology measures narrow-line fluxes. Second, low-mass halos produce stronger perturbations to more compact sources, and the dusty torus around the background quasar emits in the mid-IR from a region roughly order of magnitude more compact than the nuclear narrow-line region. 
	
	To interpret the joint constraint on $\sigma_0$ and $v_0$, we recast the inference in terms of the cross section at $20 \ \rm{km} \ \rm{s^{-1}}$, or $\sigma_{20} \equiv \sigma\left(\sigma_0, v = 20\right)$ assuming a uniform prior on $\sigma_{20}$. First, we sample the $\left(\sigma_0, v_0\right)$ prior used for the lensing analysis, compute $\sigma_{20}$ for each sample, and re-bin to derive the effective prior on $\sigma_{20}$ that corresponds to the uniform prior on $\sigma_0$ and $v_0$. This effective prior is shown in light gray in the upper panel of Figure \ref{fig:priorposteriorlikelihoodcdm}. We then repeat the sampling while weighting each $\sigma_0$ and $v_0$ by the joint probability distribution shown in Figure \ref{fig:exampleconstraintcdm} to obtain the posterior distribution from lensing. To isolate the likelihood $\mathcal{L}\left(\sigma_{20} | \rm{data}\right)$ (where data corresponds to the lensing observables), we divide the posterior by the effective prior. The resulting probability distribution $p\left(\sigma_{20} | \rm{data}\right)$ shown in the lower panel of Figure \ref{fig:priorposteriorlikelihoodcdm} has the interpretation of a posterior distribution for $\sigma_{20}$, assuming a uniform prior on $\sigma_{20}$. 
	
	The lower panel of Figure \ref{fig:priorposteriorlikelihoodcdm} demonstrates that lensing can rule out large amplitude cross sections. A larger sample of lenses, particularly with flux ratios measured in the mid-IR, can further improve the constraints, as shown by the forecast in Figure \ref{fig:forecast}. The curves show the upper limit on $\sigma_{20}$ at $95 \% \ \rm{CI}$ as a function of the number of lenses and the type of flux ratio measurement. We compute the solid curves by bootstrapping the sample of 30 lenses, applying the same procedure to compute $p\left(\sigma_{20} | \rm{data}\right)$ as described in the previous paragraph, and extrapolate (dashed curves) to estimate the constraints possible with 50 quads\footnote{Due to the computational expense of the simulations we perform, we extrapolate from 30 lenses rather than simulate a full set of 50.}. The lower (upper) bounds correspond to half (one quarter) tidal disruption efficiency in elliptical galaxies relative to the Milky Way \citep{Nadler++21}. Lensing places tighter constraints on the cross section with higher subhalo mass function amplitudes because the lensing signal scales with the number of core collapsed subhalos due to their efficient lensing properties (Figure \ref{fig:magnification}). Assuming a prior on $\Sigma_{\rm{sub}}$ of $\mathcal{N}\left(0.05, 0.01\right)$, which corresponds to doubly efficient tidal stripping in the Milky Way compared to massive ellipticals, a sample of 50 lenses can rule out cross sections $\sigma_{20} > 11 \ \rm{cm^2}\rm{g^{-1}}$ at 95$\%$CI. Assuming a prior of $\mathcal{N}\left(0.05, 0.01\right)$ on $\Sigma_{\rm{sub}}$, corresponding to $25\%$ more efficient tidal stripping in the Milky Way than in ellipticals, the constraint is $\sigma_{20} < 23 \ \rm{cm^2}\rm{g^{-1}}$ at $95 \%$CI. A sample of 30 lenses with mid-IR flux measurements that we will obtain through JWST GO-02046 (PI Nierenberg) can rule out cross sections with $\sigma_{20}$ greater than $22-28 \ \rm{cm^2}\rm{g^{-1}}$, depending on the amplitude of the subhalo mass function. The constraints with mid-IR data surpass those attainable with the narrow-line data due to the more precise measurements of the mid-IR fluxes, and the more compact background source, which increases sensitivity to low-mass halos. 
	
	\subsection{Prospects for ruling out CDM}
	\label{ssec:constraintssidm}
\begin{figure}
		\includegraphics[clip,trim=0cm 0cm 0cm
		0cm,width=.48\textwidth,keepaspectratio]{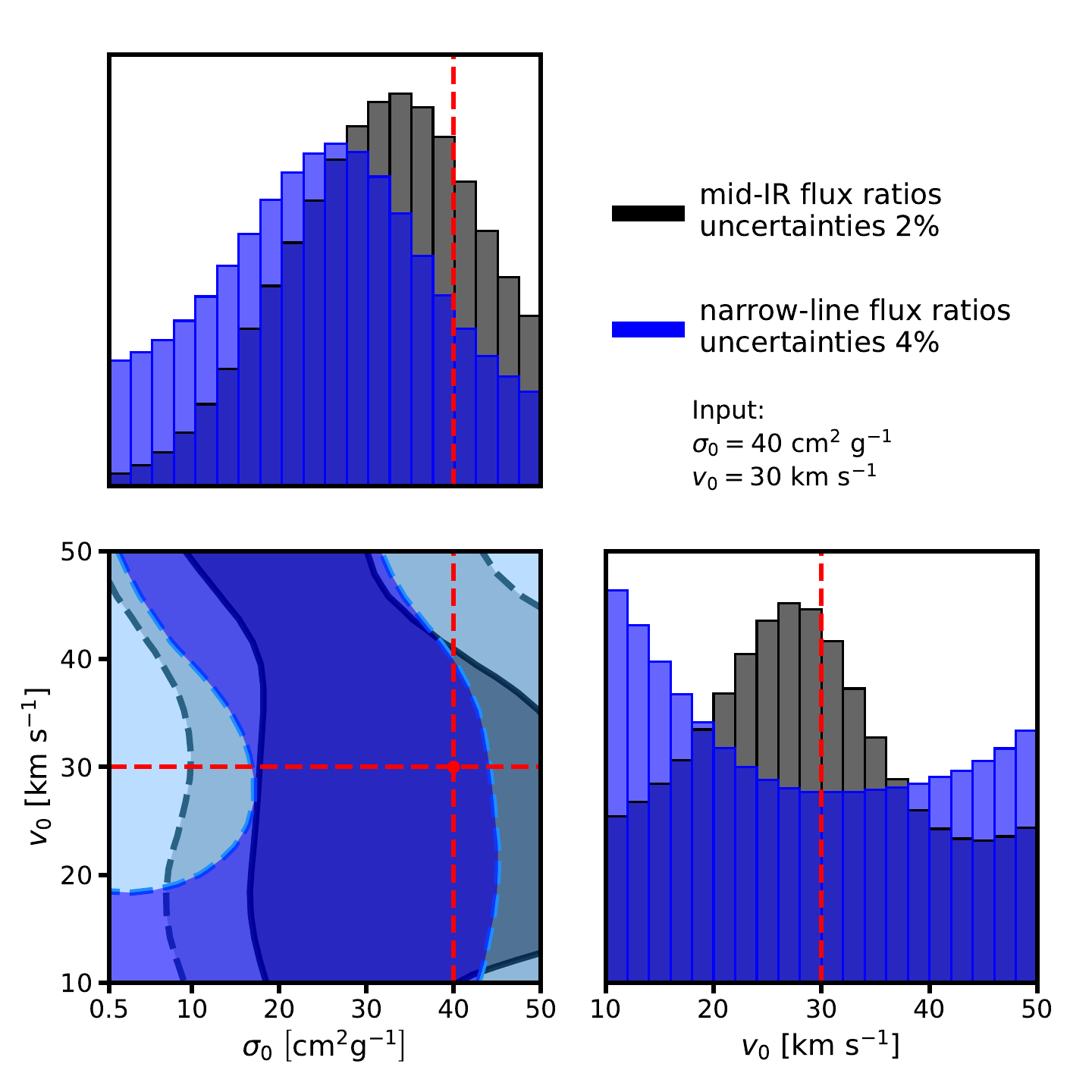}
		\caption{\label{fig:exampleconstraintsidm} Joint inference on $\sigma_0$ and $v_0$ from 30 mock lenses created with $\sigma_0 = 40 \rm{cm^2}\rm{g^{-1}}$ and $v_0 = 30 \rm{km}\rm{s^{-1}}$, or a cross section at $30 \ \rm{km} \ \rm{s^{-1}}$ of $10 \ \rm{cm^2}\rm{g^{-1}}$. The color scheme is the same as in Figure \ref{fig:exampleconstraintcdm}.}
	\end{figure}

\begin{figure}
		\includegraphics[clip,trim=0cm 0cm 0cm
		0cm,width=.45\textwidth,keepaspectratio]{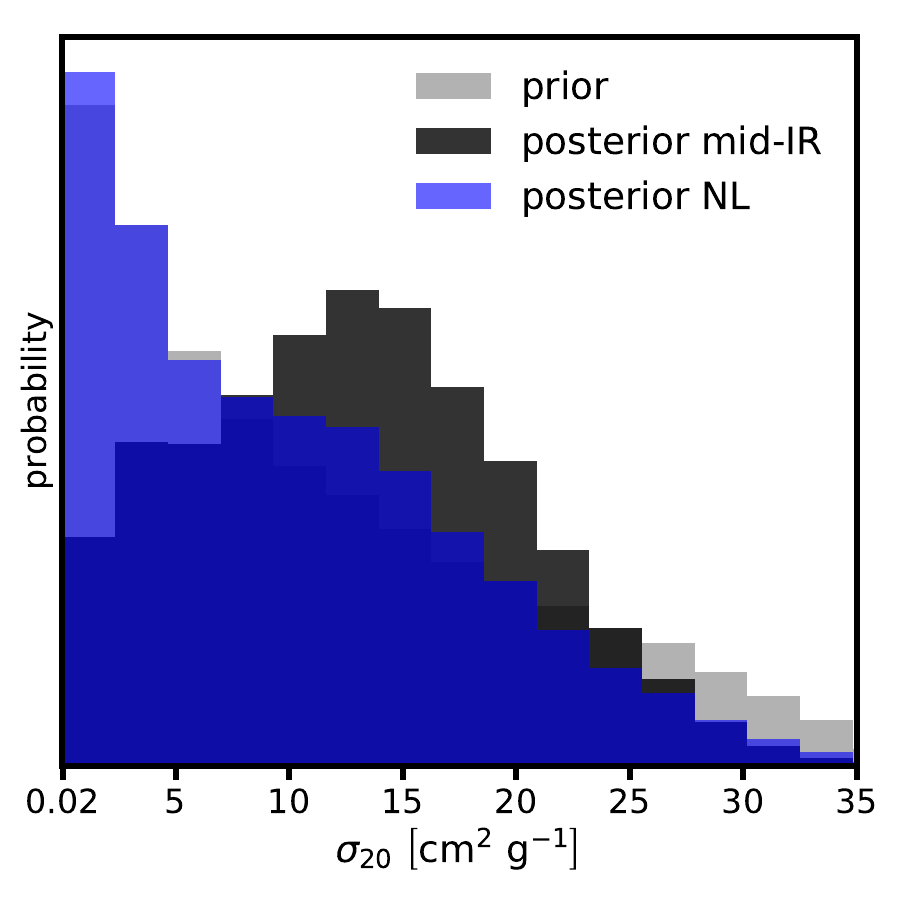}
		\includegraphics[clip,trim=0cm 0cm 0cm
		0cm,width=.45\textwidth,keepaspectratio]{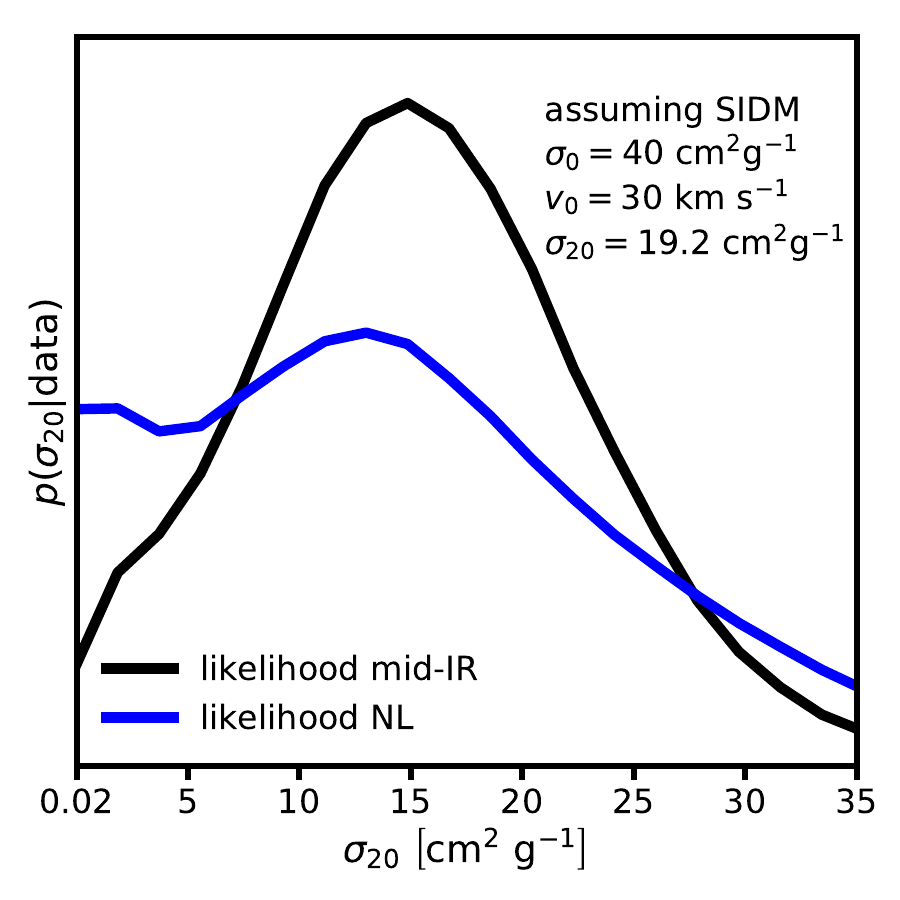}
		\caption{\label{fig:priorposteriorlikelihoodsidm} {\bf{Top:}} The effective prior and posterior distributions from Figure \ref{fig:exampleconstraintsidm} for the cross section at $20 \rm{km} \ \rm{s^{-1}}$ $\left(\sigma_{20}\right)$. The color scheme is the same as in Figure \ref{fig:priorposteriorlikelihoodcdm}. {\bf{Bottom:}} The likelihood of $\sigma_{20}$ from the lensing analysis, or the posterior distribution $p\left(\sigma_{20} | \rm{data}\right)$ assuming a uniform prior on $\sigma_{20}$, computed by sampling the posterior distribution in Figure \ref{fig:exampleconstraintsidm} and dividing out the prior.}
	\end{figure}

	We consider an SIDM model with $\sigma_0 = 40 \ \rm{cm^2} \rm{g^{-1}}$ and $v_0 = 30 \ \rm{km} \ \rm{s^{-1}}$, which corresponds to a cross section $\sigma_{20} = 19.2 \ \rm{cm^2} \rm{g^{-1}}$ and an amplitude on dwarf galaxy scales $v \sim 30 \ \rm{km} \ \rm{s^{-1}}$ of $10 \rm{cm^2} \rm{g^{-1}}$ assuming a prior on the normalization of the subhalo mass $\mathcal{N}\left(0.05, 0.01\right)$. A cross section of this amplitude may explain some of the diverse features of dwarf galaxy rotation curves in the local volume \citep{Kahlhoefer++19}. Figure \ref{fig:exampleconstraintsidm} shows the joint constraints on $\sigma_0$ and $v_0$ from a sample of 30 mock lenses with mid-IR and narrow line flux ratios. We map constraints in this two dimensional parameter space onto a constraint on $\sigma_{20}$ using the same procedure described in the previous section, and show the results in Figure \ref{fig:priorposteriorlikelihoodsidm}. While neither the sample of 30 narrow-line lenses or mid-IR lenses can decisively rule out CDM, for this particular sample of 30 lenses the mid-IR data favors SIDM over CDM with a relative likelihood of approximately $4:1$. When quoting likelihood ratios, we use the ratio of the peak of the postreior distribution to the probability of obtaining $\sigma_{20} < 2 \ \rm{cm^2}\rm{g^{-1}}$.  
	
\begin{figure}
		\includegraphics[clip,trim=0cm 0cm 0cm
		0cm,width=.48\textwidth,keepaspectratio]{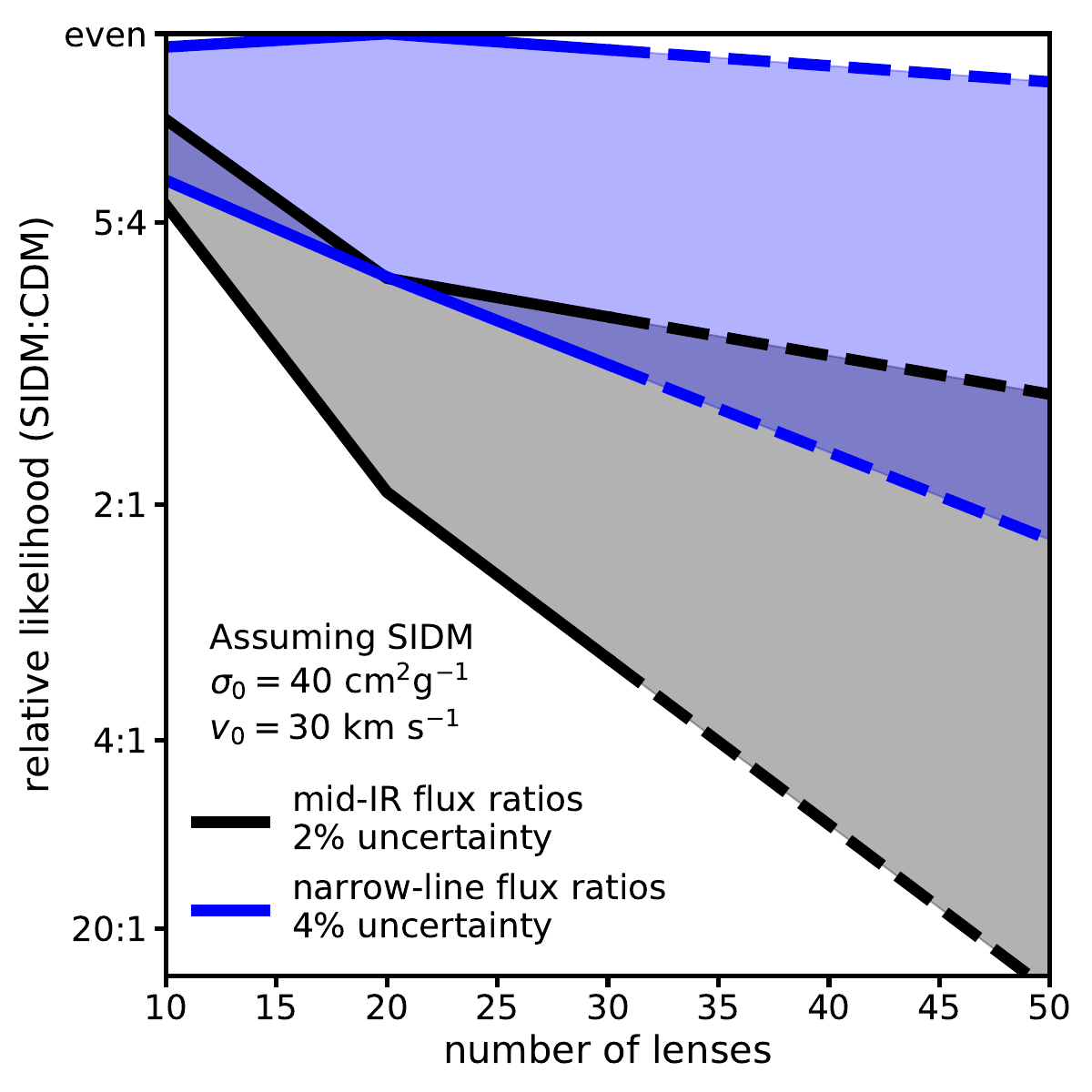}
		\caption{\label{fig:forecastsidm} Forecasts for the relative likelihood of SIDM to CDM, defined as the ratio of the most probable value of the likelihood $p\left(\sigma_{20} | \rm{data}\right)$ to the probability that $\sigma_{20} < 2 \ \rm{cm^2} \rm{g^2}$, computed as a function of the number of lenses and the type of flux ratio measurement through bootstrapping. Like Figure \ref{fig:forecast}, the lower and upper ranges of the forecast encompass the range of theoretical uncertainty associated with the amplitude of the subhalo mass function, which we implement through priors on $\Sigma_{\rm{sub}}$ of $\mathcal{N}\left(0.05, 0.01\right)$, and $\mathcal{N}\left(0.032, 0.007\right)$, respectively. The uncertainties of the flux ratio measurements indicate the expected or current precision for each dataset (see caption in Figure \ref{fig:forecast}). The input model used to create the mock datasets for the forecast has $\sigma_0 = 40 \ \rm{cm^2}\rm{g^{-1}}$ and $v_0 = 30 \ \rm{km} \ \rm{s^{-1}}$, corresponding to $\sigma_{20} = 19.2 \ \rm{cm^2}\rm{g^{-1}}$ and a cross section at $30 \ \rm{km} \ \rm{s^{-1}}$ of $10 \ \rm{cm^2}\rm{g^{-1}}$.}
	\end{figure}

	Figure \ref{fig:forecastsidm} forecasts the confidence with which a strong lens dataset can rule out CDM, assuming an SIDM model with $\sigma_0 = 40 \ \rm{cm^2} \rm{g^{-1}}$ and $v_0 = 30 \ \rm{km} \ \rm{s^{-1}}$ to CDM. The upper and lower bounds of each shaded region correspond to priors on the amplitude of the subhalo mass function of $\mathcal{N}\left(0.032, 0.007\right)$ and $\mathcal{N}$ $\left(0.05, 0.01\right)$, respectively. Similar to the trend in Figure \ref{fig:forecast}, models with more subhalos admit stronger constraints on the cross section, likely due to the increased number of core collapsed subhalos and their high lensing efficiency. For the SIDM cross section assumed in the forecast, only the mid-IR dataset reaches a statistically significant relative likelihood of SIDM to CDM of 20:1 with a sample size of 50 lenses. For a sample of 30 lenses attainable in the near future, the constraints are significantly weaker, with a relative likelihood of between 3:1 and 4:1, on average. 	
	
	\section{Discussion and conclusions}
	\label{sec:conclusions}
	We have developed a framework to use a sample of quadruple image strong gravitational lenses to constrain the interaction cross section of self-interacting dark matter in halo masses between $10^6 - 10^{10}\msun$, or velocity scales below $30 \ \rm{km} \ \rm{s^{-1}}$, accounting for both core formation and collapse in subhalos and field halos. We use a hierarchical Bayesian inference framework to validate lensing as a probe of SIDM, and to forecast constraints on $\sigma_{20}$, the cross section at $20 \ \rm{km} \ \rm{s^{-1}}$, a velocity scale where current techniques have not yet placed constraints on the cross section. We quote constraints as a function of the number of lenses, the size of the lensed background source, and the amplitude of the subhalo mass function. Our forecasts account for uncertainty in amplitude of the line of sight halo mass function, the logarithmic slope of the subhalo mass function, the background source size, and the mass profile of the main deflector, including boxyness and diskyness. We summarize our main results as follows:  
	
	\begin{itemize}
		\item A sample of 50 quads with image fluxes measured in the mid-IR at $2\%$ precision can rule out interaction cross sections greater than $11 \ \rm{cm^2}\rm{g^{-1}}$ at a relative velocity of $20 \ \rm{km} \ \rm{s^{-1}}$ at $95 \% \ \rm{CI}$, assuming a prior on the subhalo mass function that assumes tidal stripping in elliptical galaxies is half as efficient as in the Milky Way. The constraints scale inversely with the amplitude of the subhalo mass function. Assuming tidal disruption in the Milky Way is only $25 \%$ more efficient, we obtain $\sigma_{20} < 23 \ \rm{cm^2}\rm{g^{-1}}$ at $95\%$CI. A sample size of 31 lenses with mid-IR fluxes that we will obtain through JWST GO-02046 (PI Nierenberg) will rule out $\sigma_{20}$ of $22-28 \ \rm{cm^2}\rm{g^{-1}}$, assuming CDM. The constraints obtained from the mid-IR datasets improve on the constraints attainable with an equally large sample of narrow-line lenses due to both the more compact background source of the mid-IR emission, and the measurement precision attainable with JWST.
		\item A sample of 50 quads with mid-IR flux ratios can disfavor CDM with a relative likelihood of $20:1$ if dark matter has an interaction cross section at $20 \ \rm{km} \ \rm{s^{-1}}$ of $19.2 \ \rm{cm^2}\rm{g^{-1}}$, corresponding to $\sigma_0 = 40 \ \rm{cm^2}\rm{g^{-1}}$ and $v_0 = 30 \ \rm{km} \ \rm{s^{-1}}$, assuming the functional form for the interaction cross section given by Equation \ref{eqn:crossec}, and assuming tidal stripping in elliptical galaxies is half as efficient as in the Milky Way. This constraint weakens to a relative likelihood of only $3:2$ assuming an amplitude for the subhalo mass function resulting from tidal stripping that is $25 \%$ more efficient in the Milky Way than in ellipticals. While weaker than the constraints from mid-IR data, narrow-line datasets can deliver constraints with relative likelihoods of $2:1$, indicating that a larger sample of narrow-line lenses is required to place stringent constraints on the cross section. 
	\end{itemize}
	The dependence of our forecasts on the background source size highlights both the importance of accounting for finite-source effects when interpreting flux ratio statistics, and the utility of obtaining flux ratio measurements from the spatially compact mid-IR emission region around the lensed quasar. In the coming years, observations of lensed quasars with both HST and JWST will deliver flux measurements from both mid-IR and narrow-line emission for certain systems. By exploiting the differential magnification of finite size sources, an inference based on measurements from two source sizes could lead to stronger constraints than considering fluxes measured from the mid-IR or narrow-line data alone. 
	
	As a standalone method, substructure lensing provides new avenues to test the predictions of SIDM. In contrast to existing approaches, a strong lensing analysis with flux ratios probes halo mass profiles over a redshifts $z = 0 - 3$, and therefore encodes the temporal evolution of SIDM density profiles over several Gyr of time. As a purely gravitational phenomenon, lensing does not rely on baryons to infer the halo mass profile, enabling inferences of halo density profiles below $10^8 \msun$, where halos contain little stars or gas. The unique capabilities of lensing as a probe of SIDM compliment existing methods of testing the predictions of SIDM in the local volume, motivating a joint analysis \citep[e.g.][]{Nadler++21} with methods that analyze dwarf galaxies or stellar streams \citep[e.g.][]{Nadler++21b,Banik++21}. 
	
	The dependence of our forecasts on the amplitude of the subhalo mass function demonstrates that constraints on the cross section vary proportionally with the number of core collapsed subhalos. This trend supports the interpretation that the population of core collapse subhalos dominates the lensing signal, and inferences on SIDM models from lensing therefore depend primarily on the accuracy of the model used to implement core collapse. Three particular aspects of the core collapse model we implement warrant further study. First, follow up investigation should verify the accuracy of Equation \ref{eqn:collapsetimescale} for the characteristic timescale $t_0$ of the structural evolution of SIDM halos, specifically in the context of predicting the onset of core collapse as a function of halo mass and the cross section amplitude. Second, targeted simulations should examine the dependence of the distribution of collapse times around $t_0$ on the structural and orbital properties of subhalos embedded in the tidal field of a massive elliptical galaxy. Third, simulations of structure formation should test the accuracy of Equation \ref{eqn:collapseprofile} as a model for the final density profile of core collapsed halos. Assuming the functional form for the profile in Equation \ref{eqn:collapseprofile}, the relevant physical quantities include the logarithmic profile slope, the core radius, and the radius at which the core collapsed halo encloses the same mass that the NFW profile would have had in CDM. Both the core collapsed halo density profile and the timescale for core collapse determine the constraining power of lensing over SIDM models, such that these aspects of the model are somewhat degenerate in terms of our forecasts. In Appendix  \ref{sec:appA}, we illustrate how different modeling choices for the collapsed density profile affect the lensing efficiency per halo, quantified in terms of the magnification cross section. We expect to achieve similar constraints for a population of fewer core collapsed halos if they become more efficient lenses. If we fix the collapsed halo density profile, we can gauge the dependence of our forecasts on the core collapse timescale by assuming the constraints vary more or less proportionally with the number of collapsed halos, although we stress that this likely oversimplifies the problem. For example, if we had assumed a longer timescale for core collapse of $50 t_0$ rather than $10 t_0$, five times fewer, or only $\sim 5\%$, of a subhalos in the mass range $10^6 - 10^{10} \msun$ would experience core collapse, and the population of cored halos would likely have lensing properties very similar to CDM, as suggested by Figure \ref{fig:magnification}. Qualitatively, we expect thus that more lenses will be necessary to constrain the cross section parameters at the same level of precision if the timescale is longer than our assumed baseline, and viceversa. Future theoretical work constraining specifically the timescale fore collapse would be particularly helpful in case of a detection, or limits from observational data.
	
	Globular clusters can have high central densities, reminiscent of core collapsed objects, and could conceivably contribute a source of systematic uncertainty in the modeling \citep[e.g.][]{He++20}. However, the number density of subhalos in the mass range $10^6-10^7 \msun$ exceeds the median surface density of globular clusters of $4 \  \rm{arcsec^{-2}}$ quoted by \citet{He++20} by at least a factor of five, assuming an amplitude of the subhalo mass function $\Sigma_{\rm{sub}} = 0.025 \ \rm{kpc^{-2}}$, a halo mass of $10^{13.3}\msun$, a logarithmic slope $\alpha=-1.9$, and a lens redshift $z=0.5$, using Equation \ref{eqn:subhalomfunc}. The finite background source size would likely wash out the signal from globular clusters with mass $<10^6 \msun$. Regardless, we can directly address this potential systematic using our inference pipeline by including a population of globular clusters in the forward model. 
	
	The forecasts we present account for uncertainty related to the main deflector mass profile in the form of boxyness and diskyness. We implement boxyness and diskyness by adding an octopole mass moment aligned with the position angle of the elliptical mass profile. The inclusion of boxyness and diskyness more accurately describes the projected mass profiles of elliptical galaxies, and mitigates possible sources of systematic uncertainty connected to the assumed mass model of the main deflector \citep{Gilman++17,Hsueh++18}. We marginalize over the amplitude of the octopole moment in our forecasts to maintain the same degree of sophistication between the forecasts we present in this work and updated constraints on warm and cold dark matter using real datasets (Gilman et al., in prep). Using the real datasets, we will assess the impact of the more flexible mass profile on lensing constraints. 
	
	We can extend the framework presented in this work to any particle physics model that predicts $\sigma\left(v\right)$, as our model only requires knowledge of the thermal average $\langle \sigma \left(v\right) v \rangle$ in order to predict halo density profiles. To give a specific example, resonances in the cross section that boost $\langle \sigma\left(v\right) v \rangle$ by as much as a factor of ten on velocity scales $<10 \ \rm{km} \ \rm{s^{-1}}$ \citep[e.g.][]{Tulin++13} could cause a large fraction of subhalos, and even some field halos, to core collapse. The paucity of luminous matter inside halos in this mass range is of no consequence for a strong lensing analysis. 
	
	\section*{Acknowledgments}
	We thank Xiaolong Du, Manoj Kaplinghat, Ethan Nadler, Annika Peter, and Carton Zeng for useful discussions. In particular, we thank Alex Kusenko for helpful explanations and illustrations of how different particle physics models manifest in the form of the interaction cross section. We also thank the anonymous referee for constructive feedback.
	
	DG was partially supported by a HQP grant from the McDonald Institute (reference number HQP 2019-4-2). DG and JB acknowledge financial support from NSERC (funding reference number RGPIN-2020-04712). We acknowledge support from the NSF through grant NSF-AST-1714953 ``Collaborative Research: Investigating the nature of dark matter with gravitational lensing" and from NASA through grants HST-GO-15320, HST-GO-15632, HST-GO-15177. Support for this work was also provided by the National Science Foundation through NSF AST-1716527. AB acknowledges support from NASA award number 80NSSC18K1014. 
	
	This work consumed approximately 800,000 CPU hours distributed across three computing clusters. First, we performed computations on the Niagara supercomputer at the SciNet HPC Consortium \citep{Loken++10,Ponce++19}. SciNet is funded by: the Canada Foundation for Innovation; the Government of Ontario; Ontario Research Fund - Research Excellence; and the University of Toronto. Second, we used computational and storage services associated with the Hoffman2 Shared Cluster provided by the UCLA Institute for Digital Research and Education's Research Technology Group. Third, we used the {\tt memex} compute cluster, a resource provided by the Carnegie Institution for Science. 
	
	\section*{Data Availability}
	The data underlying this article will be shared on reasonable request to the corresponding author.
	
	\bibliographystyle{mnras}
	\bibliography{SIDM_bib}
	
	\appendix
	
	\section{Magnification cross section with different core collapse modeling assumptions}
	\label{sec:appA}
	Our simulations suggest that the constraining power of flux ratio statistics over SIDM models stems from the overall number of collapsed halos, and their density profiles. We expect that a higher lensing efficiency per halo can offset a smaller number of core collapsed profiles, effectively leading to a covariance between parameters that characterize the profile and the core collapse timescale, which determine the lensing efficiency and the overall number of collapsed objects, respectively. 
			
\begin{figure*}
		\includegraphics[clip,trim=0cm 0cm 0cm
		0cm,width=.48\textwidth,keepaspectratio]{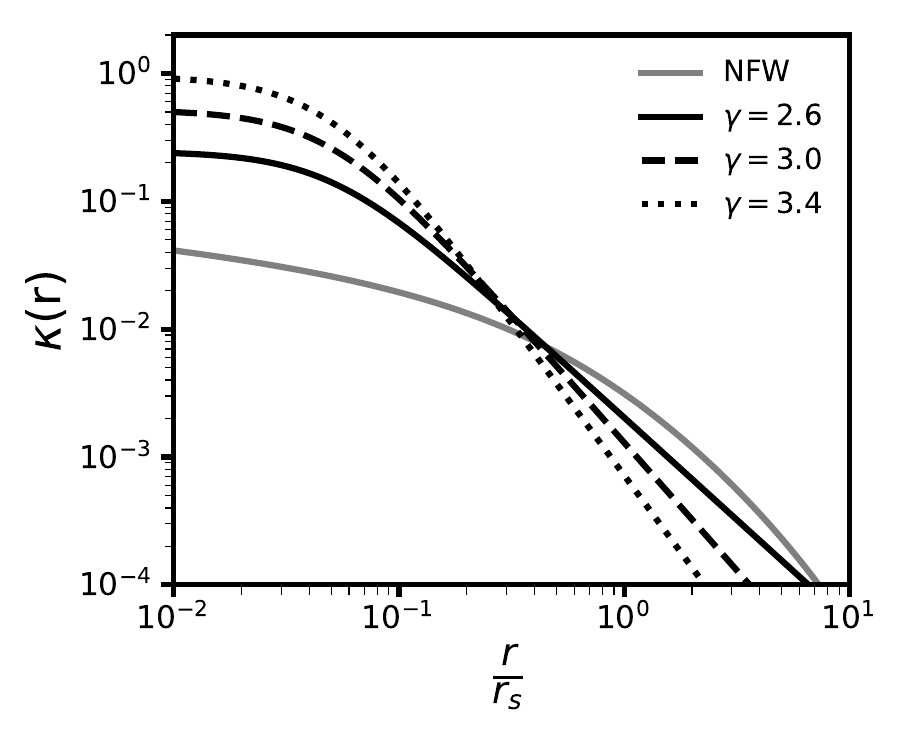}
		\includegraphics[clip,trim=0cm 0cm 0cm
		0cm,width=.48\textwidth,keepaspectratio]{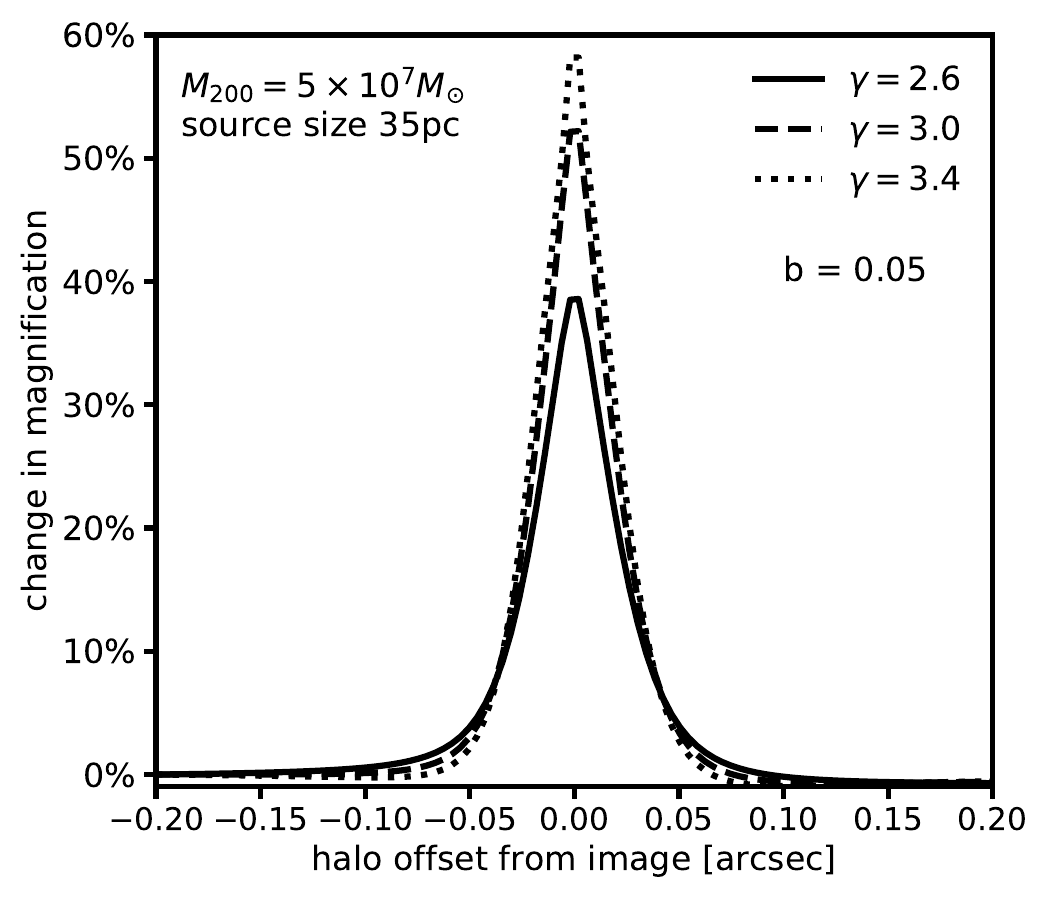}
		\caption{\label{fig:testprofileslope} The effect of changing the logarithmic slope of the core collapsed profile on the magnification cross section. The left panel shows the collapsed halo density profiles as a function of the logarithmic slope $\gamma$, matching the enclosed mass of the NFW profile within $r_{\rm{match}} = r_{\rm{vmax}} = 2.16 r_s$. The right panel shows the magnification cross section for the three different core collapsed profiles. Increasing the logarithmic slope creates more efficient lenses at small projected separations between a halo and a lensed image, but the effect has a shorter range.}
	\end{figure*}

\begin{figure*}
		\includegraphics[clip,trim=0cm 0cm 0cm
		0cm,width=.48\textwidth,keepaspectratio]{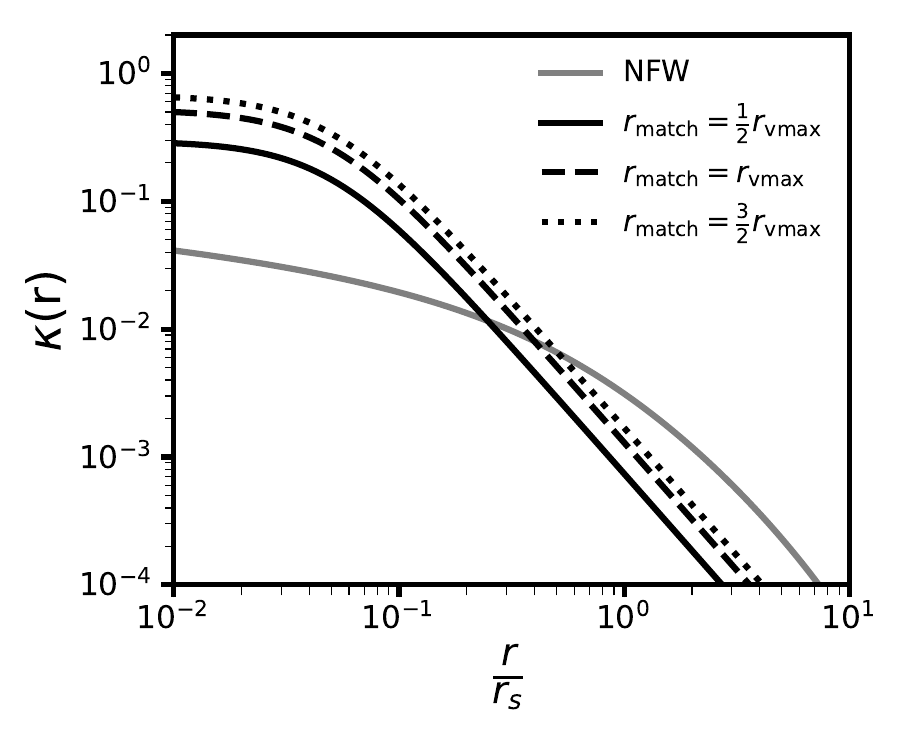}
		\includegraphics[clip,trim=0cm 0cm 0cm
		0cm,width=.48\textwidth,keepaspectratio]{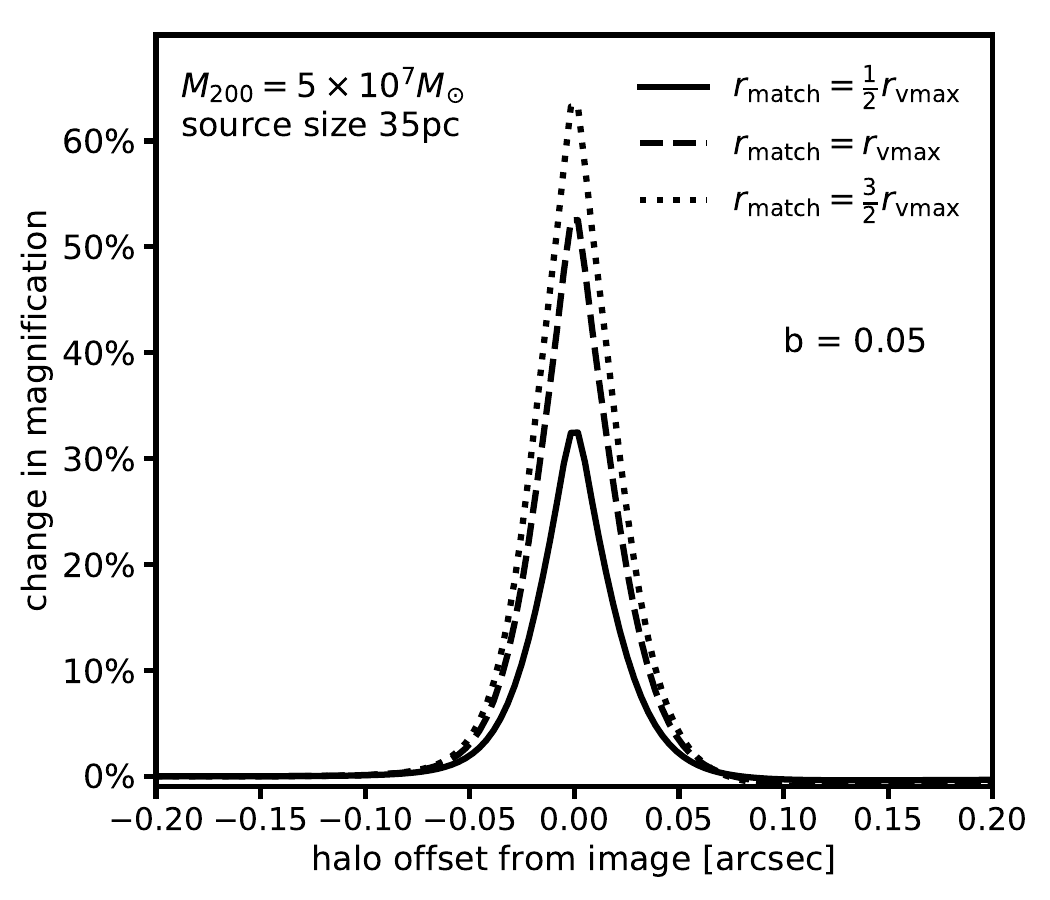}
		\caption{\label{fig:testxmatch} The effect on the magnification cross section of changing the radius at which the core collapsed profile encloses the same mass as an NFW profile. The left panel shows the projected halo density profile as a function of the matching radius $r_{\rm{match}}$, assuming a logarithmic slope $\gamma = -3$. The right panel shows the magnification cross section for each profile. Because the NFW profile drops less steeply than the collapsed profile inside $r_s$ with $\gamma = -3$, increasing $r_{\rm{match}}$ increases the mass of core collapse profile, and hence increases the magnification cross section.}
	\end{figure*}

			We identify two modeling choices that affect the lensing efficiency of core collapsed profiles in our model. Figure \ref{fig:testprofileslope} illustrates how changing the logarithmic slope $\gamma$ of the halo between 2.6 and 3.4 affects the magnification cross section. This is a larger variation in $\gamma$ that we considered in our forecasts. A steeper slope produces a more efficient lens when the halo lies near an image in projection, but the magnification perturbation has a shorter range. Figure \ref{fig:testxmatch} shows how changing the mass at which the core collapsed profile encloses the same mass as an NFW profile affects the magnification cross section. This modeling choice effectively points to a halo mass definition problem for core collapsed objects that targeted simulations can help resolve. 
			
			The effect of the halo density profile on the magnification cross sections clearly demonstrates that the number of lenses required to constrain the cross section depends on how one constructs the profiles. Specifically, because the mass definition for the core collapsed halos can significantly boost or reduce the lensing signal per collapsed halo, we would expect stronger (weaker) constraints when conserving mass between the collapsed and NFW profiles within $\frac{3}{2} r_{\rm{vmax}}$ ($\frac{1}{2} r_{\rm{vmax}}$). The prescription for assigning masses to core collapsed halos, given the mass of the halo at infall, depends on tidal disruption from the host halo, scattering between particles in the subhalo and host, and how much mass the halo ejects during contraction. 
			
			Finally, as we illustrate in Figure \ref{fig:SIDMkappa}, some collapsed profiles achieve super-critical central densities, and can produce multiple images. A higher normalization for core collapsed profiles would result in a larger fraction of subhalos capable of acting as strong lenses themselves. These objects could potentially produce distortions in lensed arcs visible by eye.
	
\end{document}